\documentclass[aps,twocolumn,preprintnumbers,amsmath,amssymb]{revtex4}
\usepackage{amsmath,amssymb,dsfont}
\usepackage{graphics,epsfig,mathtools}
\usepackage{graphicx}
\usepackage{color}
\usepackage{hyperref}
\usepackage{mathtools}
\usepackage{esint}
\usepackage{xcolor}

\def\build#1_#2^#3{\mathrel{\mathop{\kern 0pt#1}\limits_{#2}^{#3}}}

\renewcommand{\tilde}{\widetilde}          % wider `tilde'
\DeclareMathSymbol{\leqslant}{\mathalpha}{AMSa}{"36} % nicer `smaller or equal'
\DeclareMathSymbol{\geqslant}{\mathalpha}{AMSa}{"3E} % nicer `larger or equal'
\DeclareMathSymbol{\eset}{\mathalpha}{AMSb}{"3F}     % nicer `emptyset'
\renewcommand{\leq}{\;\leqslant\;}                   % redef. of < or =
\def\build#1_#2^#3{\mathrel{\mathop{\kern 0pt#1}\limits_{#2}^{#3}}}

\newcommand{\R}{\mathbb{R}}

\newcommand{\K}{\mathcal{K}}

\def \E{ \mathbb E  }
\begin{document}

\title{Numerical simulations of  a stochastic dynamics leading to cascades and loss of regularity:  applications to fluid turbulence and generation of fractional Gaussian fields}
%\\ . \\ \today}

\author{Geoffrey Beck$^a$, Charles-Edouard Br\'ehier$^b$, Laurent Chevillard$^{c,d}$, Ricardo Grande$^e$, Wandrille Ruffenach$^c$}
\affiliation{
$^{\text{a}}$Univ Rennes, IRMAR UMR 6625 \& Centre Inria de l'Universit\'e de Rennes (MINGuS) \& ENS Rennes,
France\\
$^{\text{b}}$Universite de Pau et des Pays de l’Adour, E2S UPPA, CNRS, LMAP, Pau, France\\
$^{\text{c}}$Univ Lyon, Ens de Lyon, Univ Claude Bernard, CNRS, Laboratoire de Physique, 46 all\'ee d’Italie F-69342 Lyon, France\\
$^{\text{d}}$CNRS, ICJ UMR5208, Ecole Centrale de Lyon, INSA Lyon, Universite Claude Bernard Lyon 1, Université Jean Monnet, 69622 Villeurbanne, France.\\
$^{\text{e}}$ School for Advanced Studies (SISSA), via Bonomea 265, 34136 Trieste, Italy
}%

\begin{abstract}
Motivated by the modeling of the spatial structure of the velocity field of three-dimensional turbulent flows, and the phenomenology of cascade phenomena, a linear dynamics has been recently proposed \cite{ApoBec23} able to generate high velocity gradients from a smooth-in-space forcing term. It is based on a linear Partial Differential Equation (PDE) stirred by an additive random forcing term which is delta-correlated in time. The underlying proposed deterministic mechanism corresponds to a transport in Fourier space which aims at transferring energy injected at large scales towards small scales. The key role of the random forcing is to realize these transfers in a statistically homogeneous way. Whereas at finite times and positive viscosity the solutions are smooth, a loss of regularity is observed for the statistically stationary state in the inviscid limit. We here present novel simulations, based on finite volume methods in the Fourier domain and a splitting method in time, which are more accurate than the pseudo-spectral simulations from~\cite{ApoBec23}. We show that the novel algorithm is able to reproduce accurately the expected local and statistical structure of the predicted solutions. We conduct numerical simulations in one, two and three spatial dimensions, and we display the solutions both in physical and Fourier spaces. We additionally display key statistical quantities such as second-order structure functions and power spectral densities at various viscosities.
\end{abstract}

%\pacs{02.50.Fz, 47.53.+n, 47.27.Gs}

\maketitle

\section{Introduction}

\subsection{The numerical investigation of a stochastic transport equation in Fourier space}

The purpose of this article is the numerical simulation of a recently proposed model of fully developed fluid turbulence \cite{ApoBec23}. This model of a new kind is based on a temporal evolution, governed by a linear partial differential equation (PDE), randomly stirred by an additive force which is smooth and homogeneous in space and delta-correlated in time. Similar additive type of forcing has been traditionally used for numerical investigations of Navier-Stokes (NS) equations in order to observe the turbulent behavior of the velocity field \cite{Fri95,Pop00}. As explained in the sequel, the formulation of the aforementioned model for the dynamics is especially convenient in Fourier space. The proposed underlying mechanism, that is able to transfer energy as a turbulent cascade would do, is based on a transport equation in the Fourier domain, and can be seen as a simple model of the generation of small scales. These notions will be properly defined later in the article.

Preliminary numerical simulations have been performed in \cite{ApoBec23} using pseudo-spectral methods which rely heavily on the Fast Fourier Transform (FFT). Unfortunately, using necessarily Cartesian grids of these periodic boxes, spurious anisotropies were observed. In this work, we overcome this issue by applying a finite volume method in the Fourier variables formulation, on a mesh which preserves the spherical symmetry of the model, which is able to accurately describe all the elements of the dynamics (transport, damping and random forcing) and which is well-defined even when the unknows need to be understood as distributions instead of classical regular functions.

The article is organized as follows. The rest of this section is devoted to a short introduction to hydrodynamic turbulence, the presentation of the model proposed in \cite{ApoBec23} and the description of its analytical solution in a continuous formulation. The second section is devoted to the presentation of the finite-volume and splitting methods that will be applied. The formulation of this method is detailed in dimension $d=1$ for pedagogical reasons, and then special attention is paid to the more physical two and three dimensional settings with a focus on the role of polar and spherical symmetries of the finite volume mesh. In addition, we present the splitting method for the temporal discretization, where in particular a transport problem is solved exactly. In the third section, we report and discuss the numerical results. We draw conclusions and discuss possible perspectives in the last section.

\subsection{Fully developed hydrodynamic turbulence}

The phenomenology of three-dimensional fluid turbulence, although surprisingly unrelated to a wide extent to the underlying equations of motion given by the Navier-Stokes equations, is now well accepted after decades of experimental investigations and interpretations \cite{TenLum72,Fri95,Pop00}. To set ideas, consider the velocity vector field $u(t,x)$ made up of three-components $u=(u_i)_{i=1,2,3}$ at a given position $x\in\mathbb R^3$ and at time $t\ge 0$. In this article, we will focus on the simplistic situation referred to statistically stationary, homogeneous and isotropic turbulence, meaning that the probability distribution of the solution is invariant both by rotations and by spatial and temporal translations. It has been observed that this regime is reached after a transient evolution for solutions of the randomly forced Navier-Stokes equations
\begin{equation}\label{eq:NS}
\partial_t u +(u\cdot \nabla)u=-\nabla p+\nu \Delta u +f,
\end{equation}
where $u(t,x)$ is the velocity field of an incompressible, $\nu$ denotes the viscosity, $p(t,x)$ is the pressure, and $f(t,x)$ is the random forcing field. We consider a forcing localized at \textit{large scales}, {\it{i.e.}} only Fourier modes $\hat{f}(t,k)$ corresponding to wave numbers $k$ with norm $|k|$ close to $1/L$ are populated with energy \cite{Pop00}. The parameter $L$ is known as the integral length scale in turbulence literature. For a general class of random forcing terms $f(t,x)$ satisfying the condition above, it has been repeatedly observed that the asymptotic variance of velocity fluctuations converges to a positive non-trivial limit $\sigma^2 \in(0,\infty)$ in the inviscid regime $\nu\to 0$, namely
\begin{equation}\label{eq:AsymptVar}
\underset{\nu\to 0}\lim~\underset{t\to\infty}\lim~\E|u(t,x)|^2=:\sigma^2,
\end{equation}
where $\E$ stands for the mathematical expectation with respect to the instances of the random forcing $f$. In \eqref{eq:AsymptVar}, the asymptotic velocity variance $\sigma^2$ is a positive and finite number, which depends, for instance, in an intricate way, on the boundary conditions, if any, and on the details of the forcing field $f$. Moreover, the velocity variance \eqref{eq:AsymptVar} is independent on the position $x$ as a consequence of the observed statistical homogeneity (i.e. invariance by translations of the underlying statistical laws). Thus, in order to reach \eqref{eq:AsymptVar}, turbulent fluids have to generate a mechanism which is able to dissipate in a very efficient way the energy that is injected into the system in a statistically stationary  way such that velocity fluctuations get independent of viscosity, or equivalently as the Reynolds number $\mathcal R_e\equiv \sigma L/\nu$ tends to infinity. This dissipation mechanism requires an energy transfer from large to small scales. As a result, the velocity field needs to generate small scale structures and has a \textit{rough} behavior, which can be described in several ways. In the spatial domain, this means that the solution $u(t,x)$ is only H\"older continuous with respect to $x$. In the Fourier domain, roughness can be described by the behavior of the power spectral density $E(t,k)$ (PSD), which is defined as the Fourier transform of the velocity correlation function, {\it{i.e.}} 
\begin{align}\label{eq:DefPSD}
E(t,k) = \int_{x\in\R^3}e^{-2i\pi k\cdot x}\E\left[ u(t,0)\cdot u(t,x)\right]dx.
\end{align}
The power-law decay of the PSD for large $|k|$
\begin{align}\label{eq:Kolmo41}
\underset{\nu\to 0}\lim~\underset{t\to\infty}\lim~E(t,k) \build{\propto}_{|k|\to\infty}^{}|k|^{-(2H+d)},
\end{align}
where $d$ is the spatial dimension and $H$ is the H\"older regularity exponent, is expected to hold for a general class of hydrodynamic turbulence models. For three-dimensional fully developed turbulence, with $d=3$, Kolmogorov predicted that $H\approx 1/3$ by dimensional arguments  \cite{Kol41,Fri95}. Note that the PSD depends on the norm $|k|$ of $k$ only, due to the isotropy of the model. As a result, averaging over the angle variable and weighting by the surface $4\pi|k|^2$ of the shell of radius $|k|$, the power-law decay depicted in \eqref{eq:Kolmo41} leads to the famous $|k|^{-5/3}$-law of fully developed fluid turbulence. Even if the Fourier support of the forcing field $f$ entering in \eqref{eq:NS} is limited to low wave numbers, asymptotically in the large time regime the Fourier transform $\hat{u}$ of the velocity field $u$ has a full support. To obtain this behavior and in particular the power-law decay of the PSD, the nonlinear Navier-Stokes dynamics has somehow transferred energy from the large scales towards small ones. This transport through scales is referred to the \textit{cascade} phenomenon.

The decay of the amplitude of the Fourier modes in the right-hand side of~\eqref{eq:Kolmo41} is typical of rough, non differentiable fields. However, at low wave numbers $|k|\ll 1$, the spectral energy is expected to remain finite, ensuring in particular a finite velocity correlation length in the physical space, which corresponds to, up to a viscous independent multiplicative factor, the typical length scale $L$ of the random force. Accordingly, the functional form of the power spectral density is integrable over $k$ in any space dimension $d$ when $0<H<1$, ensuring a finite variance of velocity fluctuations in physical space, in a consistent manner as required by the phenomenology of turbulence \eqref{eq:AsymptVar}. H\"older continuous but not differentiable random fields have a long history in the probabilistic modeling of turbulence \cite{Kol40,ManVan68,kraichnan1968small,
ChaGaw03,PerGar15,CheGar19,ApoChe22}. Indeed, fractional Gaussian fields are able to reproduce the second-order statistics of turbulent fluids. Let the velocity increment over a given vector $\ell\in\R^d$ be defined as
\begin{equation}\label{eq:DefIncrement}
\delta_\ell u(t,x) = u(t,x+\ell)-u(t,x),
\end{equation}
and refer to its variance as the second-order structure function in the language of the phenomenology of turbulence. As suggested by many studies, the second-order structure function behaves at small scales as
\begin{equation}\label{eq:Kolmo41SecondOrderStructFunc}
\underset{\nu\to 0}\lim~\underset{t\to\infty}\lim~\E \left[ \left| \delta_\ell u(t,x) \right|^2\right] \build{\propto}_{|\ell|\to 0}^{}|\ell|^{2H},
\end{equation}
which is consistent with \eqref{eq:Kolmo41}. The power-law behavior of the second-order structure function when $|\ell|\to 0$ is known as the $2/3$-law of turbulence \cite{Fri95} for $H=1/3$.

\subsection{A linear model of the cascade phenomenon}

\subsubsection{Presentation of the model}

To the best of our knowledge, the cascade phenomenon described above, {\it{i.e.}} the transfer of energy from the large scales towards smaller ones, has not been rigorously established for solutions of the randomly forced Navier-Stokes equations \eqref{eq:NS}. 
In particular, the property \eqref{eq:AsymptVar} that the variance of velocity fluctuations is asymptotically independent of viscosity $\nu$ in the fully developed turbulent regime ({\it{i.e.}} when the Reynolds number goes to infinity) and the Kolmogorov power-law \eqref{eq:Kolmo41} remain unexplained from first principles \eqref{eq:NS}. 
Nonetheless, very clear illustrations of this energy cascade phenomenon can be derived in a rigorous way assuming the finiteness of the velocity variance \eqref{eq:AsymptVar}, and more generally assuming the existence of a statistically stationary homogeneous and isotropic turbulent regime. 
Early formulated in terms of Onsager's conjecture \cite{Ons49}, and known in the turbulence literature in an averaged sense using the K\'arm\'an-Howarth equation and the $4/5$-law \cite{Fri95}, several mathematical developments and arguments have been proposed \cite{Eyi94,ConE94,DeLSze14,BucVic20}, see for instance the review articles \cite{EyiSre06,Dub19}. 
Let us also mention that several of the aforementioned properties of fluid turbulence, in particular the independence of velocity variance on viscosity \eqref{eq:AsymptVar}, can be reproduced by the Burgers equation \cite{MitBec05,BorKuk21,Kuk24}, without developing a Kolmogorov spectrum \eqref{eq:Kolmo41}. 
However, those illustrations focus on higher-order statistical quantities instead of the second-order ones like the variance \eqref{eq:AsymptVar}, the correlation function \eqref{eq:DefPSD} and the second-order structure function \eqref{eq:DefIncrement}. 
The purpose of this article is to design a dedicated and effective numerical method to investigate the properties of a \textit{model}. This model is given by a linear stochastic partial differential equation (SPDE), which is much simpler than the randomly forced Navier-Stokes equations \eqref{eq:NS}. 
We will show that this model is able to reproduce the fundamental second-order statistical behavior of turbulence, described above: the property that asymptotically the velocity variance is independent of viscosity \eqref{eq:AsymptVar}, the power-law decay of the PSD \eqref{eq:Kolmo41}, and the power-law behavior of the second-order structure function \eqref{eq:Kolmo41SecondOrderStructFunc}. 
The considered model is given by a linear evolution equation driven by an additive Gaussian random forcing. Therefore the solution is a centered Gaussian random field, which is fully caracterized by studying second-order statistical quantities. 
Reproducing different scaling properties for higher-order statistical quantities is not possible for the model considered in this work.

In the two recent articles \cite{ApoChe22,ApoChe23}, it has been proposed to model the energy cascade phenomenon via a transport equation in Fourier space. 
In particular, the authors have been able to design a SPDE for a scalar velocity field, in one-space dimension ($d=1$). In the simplest situation where only a linear evolution is considered, the resulting velocity field is a Gaussian function when the random forcing is also assumed to be Gaussian, and shares many properties with those observed in fully developed turbulence, see \eqref{eq:AsymptVar}, \eqref{eq:Kolmo41} and \eqref{eq:Kolmo41SecondOrderStructFunc}. Unfortunately in \cite{ApoChe22,ApoChe23}, the velocity $u(t,x)$ and forcing $f(t,x)$ fields take complex values, and furthermore, extensions to higher spatial dimensions, in particular to $d=3$, are neither obvious nor natural. In the recent article \cite{ApoBec23}, the authors have fixed those issues and have proposed a version which provides real-valued velocity fields and in arbitrary spatial dimension $d$, which we introduce next.

The Fourier transform of a smooth function $\varphi$ which decays sufficiently fast at infinity is defined for all wave vectors $k\in\R^d$ by
\begin{equation}\label{eq:DefFT}
\widehat{\varphi}(k) = \mathcal{F} \varphi (k)= \int_{x\in\R^d}e^{-2i\pi k\cdot x}\varphi(x)dx.
\end{equation}
This definition can be generalized to the class of Schwartz tempered distributions \cite{Hor15}. Even if the same notation is employed, recall that the Fourier transform of a tempered distribution cannot be interpreted pointwise.

Let us consider an external force $f$ which is a statistically homogeneous, isotropic and stationary real-valued Gaussian field. Assume also that it is smooth in space and delta-correlated in time, and that it is centered: $\E[f(t,x)]=0$ for all $t\ge 0$ and $x\in\R^d$. The field $f$ is thus caracterized by the correlations
\begin{align}\label{eq:CorrForcingPhysSpace}
\E \left[f(t,x)f(t',x')\right] =\delta(t-t') \, C_f(x-x'),
\end{align}
where the notation $\delta(t)$ stands for the Dirac distribution and the correlation function $C_f$ is a smooth real-valued function. The radial symmetry implies that the correlation depends on $|x-x'|$ only. Considering the Fourier transform, one obtains a generalized random field $\hat{f}(t,k)$ which is centered, \textit{i.e.} one has $\E[\hat{f}(t,k)]=0$ for all $t\ge 0$ and $k\in\R^d$, and with the correlation
\begin{equation}\label{eq:CorrForcingFourSpace}
\E \left[\widehat{f}(t,k) \overline{\widehat{f}(t',k')}\right] =\delta(t-t')\delta(k-k')\widehat{C}_f(k),
\end{equation}
where $\overline{\cdot}$ denotes the complex conjugate, $\delta(k)$ stands for Dirac distribution in dimension $d$ (and is thus the product of the one-dimensional Dirac distributions for each component) and $\widehat{C}_f(k)\ge 0$ for all wave numbers $k$. Since the forcing $f$ is a real-valued field, the Fourier transform $\widehat{f}$ satisfies the Hermitian symmetry property $\overline{\widehat{f}(t,k)}=\widehat{f}(t,-k)$.

As argued above, we consider a forcing term $f$, such that its Fourier transform is compactly supported, away from the origin: there exists $\kappa>0$ and $k_f\ge \kappa$ such that
\begin{equation}\label{CFsupp}
|k| \notin ( \kappa, k_f) \implies \widehat{C}_f (k)= 0.
\end{equation}

Let $c>0$ be an additional parameter.
The stochastic partial differential equation proposed in the recent article \cite{ApoBec23} reads
\begin{align}\label{eq:DefEvolFourSpace}
\partial_t\widehat{u}+&c\left[\text{div}_k\left( \frac{k}{|k|}\widehat{u}\right) + \frac{H+\frac{1}{2}}{|k|}\widehat{u}\right] 
= -\nu (2\pi |k|)^2 \widehat{u} + \widehat{f},
\end{align}
for all $t\ge 0$ and all $k\in\R^d$ such that $|k|\ge \kappa$, where $\text{div}_k:= \sum_i \partial_{k_i}$ stands for the divergence operator with respect to $k$. In addition, the boundary condition
\begin{align}\label{eq:BCFourSpace}
\widehat{u} |_{ |k| =  \kappa } = 0
\end{align}
is imposed. Finally, for instance, the initial condition
\begin{align}\label{eq:ICFourSpace}
\widehat{u}|_{t=0 } =0,
\end{align}
is given. 

On the right-hand side of \eqref{eq:DefEvolFourSpace}, the term $-\nu (2\pi |k|)^2 \widehat{u}$ is the Fourier formulation of the viscous dissipation term $\nu\Delta u$. Observe that the linear operator appearing on the left-hand side of \eqref{eq:DefEvolFourSpace} can be decomposed as the sum of two terms
\begin{equation}\label{div-to-partial_k}
\text{div}_k\left( \frac{k}{|k|}\widehat{u}\right) = \partial_{|k|} \widehat{u} + \frac{d-1}{|k|}  \widehat{u}
\,, \quad \partial_{|k|}:= \frac{k}{|k|}\cdot \nabla_k.
\end{equation}
We identify the radial transport operator $\partial_{|k|}$ as a key ingredient in the dynamics \eqref{eq:DefEvolFourSpace}. Notice that the radial transport equation $\partial_t \widehat{u} + c\,\text{sign}(k)\partial_k \widehat{u}=0$ obtained choosing $\nu=0$, $d=1$, $H=-1/2$ and $f=0$ has the solution $\widehat{u}(t,k)=\widehat{u}(0,k-\text{sign}(k)t)$, meaning that the initial value $\widehat{u}(0,k)$ is transported to large $k\to\infty$ (resp. $k\to -\infty$) when $k>0$ (resp. $k<0$). The radial transport equation is thus able to model a cascade phenomenon.

In this work we only consider the stochastic evolution equation \eqref{eq:DefEvolFourSpace} considered in the Fourier domain. We mention that it is possible to consider an equivalent formulation in the physical space: the field $u(t,x)$ is solution of
\begin{align}\label{eq:DefEvolPhysSpace}
\partial_t u + A u = \nu \Delta u +f,
\end{align}
where the operator $A$ is formally defined by
\begin{align}\label{eq:DefAPseudoDiffPhysSpace}
A u= c \,  \mathcal{F}^{-1} \left[\text{div}_k\left( \frac{k}{|k|}\widehat{u}\right) + \frac{H+\frac{1}{2}}{|k|}\widehat{u}\right], 
\end{align}
with $\mathcal{F}^{-1}$ denoting the inverse Fourier transform. We refer to the article \cite{ApoBec23} for the rigorous definitions and analysis. The operator $A$ is the sum of a differential operator of degree 0 plus a regularizing operator. Linear equations with operators of degree 0 are also common whenever one introduces a dispersive perturbation in a hyperbolic system. In such cases, these operators of degree 0 are used to model wave propagation under strong dispersive effects and they are responsible for memory effects. See for example \cite{BecLan22} in the context of wave-energies, \cite{BecHam23} in the context of electromagnetic waves propagating along a coaxial cable and \cite{ColSai20,Col20,MaaLam95,RieVal97,BroErm16,ShaBuh23} in the context of internal or inertial waves (see also subsequent mathematical developments in Refs. \cite{DyaZwo19,GalZwo22}).

Let us now explain the role of the parameter $\kappa>0$. The linear operator appearing in the left-hand side of the evolution equation \eqref{eq:DefEvolFourSpace}, see also \eqref{div-to-partial_k}, is ill defined for $|k|=0$. To avoid this issue, like in the article \cite{ApoBec23}, it is imposed that $\widehat{u}(t,k)=0$ if $|k|\le \kappa$. This is ensured by applying the boundary conditions \eqref{eq:BCFourSpace} and by assuming also that $\widehat{f}(t,k)=0$ for $|k|\le \kappa$. As explained above and as required by the phenomenology of fluid turbulence, the random forcing is imposed only at \textit{large scales}, therefore it is also assumed that $\widehat{f}(t,k)\neq 0$ only in a shell of characteristic width $1/L$ and centered on $|k|\approx 1/L$, where $L$ has the physical meaning of the integral length scale. As soon as $1/L$ is chosen larger than $\kappa$, the behavior of the solution $\widehat{u}(t,k)$ does not depend significantly on $\kappa$.

Finally, let us comment on the additional linear term entering in the left-hand side of the evolution equation \eqref{eq:DefEvolFourSpace}, which is proportional to $|k|^{-1}$. As it is argued in the recent articles \cite{ApoChe22,ApoBec23}, considering the equation
\begin{equation}\label{eq:vg}
\partial_t\widehat{v}+c\partial_{|k|}\widehat{v}
= \widehat{g}
\end{equation}
the solution $v(t,x)$ develops a highly singular nature in space, when $t\to\infty$, close to the regularity of a white noise. In order to build rough fields which are H\"older continuous with H\"older exponent $H\in(0,1)$, an additional (linear) operation is required, based on a Fourier multiplier given by the power-law behavior $|k|^{-(H+d/2)}$. Considering the variables $\widehat{u}(t,k) = |k|^{-(H+d-1/2)} \widehat{v}(t,k)$ and $\widehat{f}(t,k)=|k|^{-(H+d-1/2)} \widehat{g}(t,k)$, from~\eqref{eq:vg} one obtains the dynamics~\eqref{eq:DefEvolFourSpace}, see the details in \cite[Section~3.2]{ApoBec23}.

\subsubsection{Formal solution and its asymptotic behavior}

The solution of the evolution equation \eqref{eq:DefEvolFourSpace}, with boundary condition provided in \eqref{eq:BCFourSpace} and with initial value given by \eqref{eq:ICFourSpace} can be expressed as follows: for all $t\ge 0$ and if $|k|\ge \kappa$, one has
\begin{align}\label{eq:MildSolution}
&\widehat{u}(t,k)= \int_{\left(t- \frac{|k| -\kappa}{c} \right)_+}^{t} e^{\frac{4\pi^2\nu}{3c} ((|k| -c(t-s))^3-|k|^3)} \notag \\
&  \left(\frac{|k|  -c(t-s)}{|k|} \right)^{H+d-\frac{1}{2}} \widehat{f} \left(s, \left(|k|  -c(t-s) \right) \frac{k}{|k|} \right)   ds,
\end{align}
where $\tau_+=\max(0,\tau)$ denotes the positive part of a real number $\tau$, see \cite[Theorem~3.7]{ApoBec23}. With the expression of the solution given in \eqref{eq:MildSolution}, one gets following correlation structure 
\begin{align} \label{eq:CorrModesSol}
    \mathbb{E} [ \widehat{u}(t,k)&\overline{\widehat{u}(t,k')} ]=\notag\\
    &\delta(k-k') |k|^{-(2H+d)} e^{-\frac{8\pi^2 \nu}{3c} |k|^3} F_\nu(t,|k|),
\end{align}
where for all $t\ge 0$ and all $k\in\R^d$ one has
\begin{align} \label{eq:FnuFiniteTime}
&F_\nu(t,|k|)= \\
  &  \begin{cases}
      \displaystyle 0     \, \text{ for }\,  |k| \leq \kappa, \\
     \displaystyle  \frac{1}{c} \int_{\kappa}^{|k|} e^{\frac{8\pi^2\nu}{3c}s^3} s^{2H+d} \widehat{C}_f(s) ds  \, \text{ for }\,  \kappa < |k| \leq ct+ \kappa,  \\
     \displaystyle    \frac{1}{c} \int_{|k|-ct}^{|k|} e^{\frac{8\pi^2\nu}{3c}s^3} s^{2H+d}  \widehat{C}_f(s) ds\, \text{ for }  \, |k| > ct+ \kappa ,
    \end{cases}
\end{align}
see \cite[(3.13) and (3.14)]{ApoBec23}.

Let us show that the properties \eqref{eq:AsymptVar} of the velocity variance, \eqref{eq:Kolmo41} of the power spectral density and \eqref{eq:Kolmo41SecondOrderStructFunc} of the second-order structure function are retrieved from \eqref{eq:FnuFiniteTime}.

First, observe that
\begin{align} \label{eq:ComputFDoubleLimit}
F(|k|) &:= \underset{\nu\to 0}\lim~\underset{t\to\infty}\lim~e^{-\frac{8\pi^2 \nu}{3c} |k|^3} F_\nu(t,|k|) \notag\\
&= \mathds{1}_{|k|\ge \kappa}\int_{\kappa}^{|k|}  s^{2H+d}  \widehat{C}_f(s) ds,
\end{align}
where $\mathds{1}_S$ denotes the indicator function of a set $S$. Indeed, for any fixed wave number $k$, with $|k|\ge \kappa$, for sufficiently large $t$ one has $|k| \leq ct+ \kappa$. Moreover, recall that the power spectral density $ \widehat{C}_f$ of the forcing is compactly supported owing to \eqref{CFsupp}, therefore the value of $F(|k|)=F(k_f)$ is independent of $|k|\ge k_f$ for large wave numbers. As a result, since the power-law function $|k|^{-(2H+d)}$ is integrable at infinity in any dimension $d\ge 1$ and for $H\in(0,1)$, one obtains
\begin{align} \label{eq:ComputVarSolDoubleLimit}
\lim_{\nu\to 0}\lim_{t\to\infty} \E u^2(t,x)=\int |k|^{-(2H+d)}F(|k|)dk\in(0,\infty),
\end{align}
see \cite[Proposition 4.10]{ApoBec23}.

Let us now study the asymptotic behavior of the PSD: for all $k$ with $|k|\ge \kappa$ one has
\begin{equation} \label{eq:ComputPSDDoubleLimit}
\underset{\nu\to 0}\lim~\underset{t\to\infty}\lim~E(t,k) = |k|^{-(2H+d)}F(|k|),
\end{equation}
where we recall that $F$ is given by \eqref{eq:ComputFDoubleLimit} and that $F(|k|)$ is independent of $|k|$ for $|k|\ge k_f$, see \cite[Proposition~4.10]{ApoBec23}.

Finally, it remains to study the second-order structure function. Owing to the power-law decay of the PSD \eqref{eq:ComputPSDDoubleLimit}, one obtains
 \begin{align} \label{eq:ComputSecondOrderSFModelDoubleLimit}
 \underset{\nu\to 0}\lim~\underset{t\to\infty}\lim~ & \E\left[\left(\delta_\ell u(t,x)\right)^2\right]\notag \\
 &= 2\int \left[ 1-\cos\left( 2\pi k\cdot \ell\right)\right]|k|^{-(2H+d)}F(|k|)dk.
\end{align}
One obtains the following when $|\ell|\to 0$, see \cite[Corollary~4.12]{ApoBec23}:
\begin{align} \label{eq:ComputSecondOrderSFModelDoubleLimitSmallScales}
 \underset{\nu\to 0}\lim~\underset{t\to\infty}\lim~&\E\left[\left(\delta_\ell u(t,x)\right)^2\right]\notag\\
 &\build{\sim}_{|\ell|\to 0}^{} 2c_d|\ell|^{2H} \int_{\kappa}^{\infty}s^{2H+d} \widehat{C}_f(s) ds,
\end{align}
which depends on the function $\widehat{C}_f$, and where $c_d$ is a geometrical factor arising from the integration in the unit sphere of dimension $d-1$ of the scalar product of $k$ with $\ell$ that enters in \eqref{eq:ComputSecondOrderSFModelDoubleLimit}. Its general expression in dimension $d$ is cumbersome, but reads explicitly in dimension $d=1$ as
\begin{align}\label{eq:PredContCdS2D1}
 c_1=2\int_{\rho=0}^\infty\left[ 1-\cos\left( 2\pi \rho\right)\right]\rho^{-(2H+1)}d\rho,
 \end{align}
in dimension $d=2$ as
 \begin{align}\label{eq:PredContCdS2D2}
 c_2=\int_{\rho=0}^\infty\int_{\theta=0}^{2\pi}\left[ 1-\cos\left( 2\pi \rho\cos \theta\right)\right]\rho^{-(2H+1)}d\rho d\theta,
 \end{align}
which could be further simplified using a Bessel function of the first kind after integration over the angular variable, and finally, in dimension $d=3$, as
 \begin{align}\label{eq:PredContCdS2D3}
 c_3=2\pi\int_{\rho=0}^\infty\int_{\theta=0}^{\pi}\left[ 1-\cos\left( 2\pi \rho\cos \theta\right)\right]\rho^{-(2H+1)}\sin \theta d\rho d\theta,
 \end{align}
which also can be simplified introducing a sine cardinal after integration over $\theta$.

Our model \eqref{eq:DefEvolFourSpace} is thus able to reproduce the second-order statistical behavior of the solutions of the forced Navier-Stokes equations, see \eqref{eq:AsymptVar}, \eqref{eq:Kolmo41} and \eqref{eq:Kolmo41SecondOrderStructFunc}.

Notice that this is not the case for the stochastic heat equation which can be seen as \eqref{eq:DefEvolFourSpace} with $c=0$, {\it{i.e.}} without the transport in $k$-space, and removing the boundary condition~\eqref{eq:BCFourSpace}. The solution of the stochastic heat equation is given by
\begin{align}\label{eq:SHEMildSolution}
\widehat{u}_{c=0}(t,k)= \int_{0}^{t} e^{-8\pi^2\nu |k|^2(t-s)}  \widehat{f} \left(s, k\right)ds.
\end{align}
and the correlation structure of the Fourier modes is given by
\begin{align}\label{eq:CorrModesSHEFourSpace}
\E \left[\widehat{u}_{c=0}(t,k) \overline{\widehat{u}_{c=0}(t,k')}\right] =\delta(k-k')\widehat{C}_f(k)\frac{1-e^{-8\pi^2\nu |k|^2t}}{8\pi^2\nu |k|^2}.
\end{align}
Then, in a similar way as we obtained the limiting behavior of the variance of the solution of our model \eqref{eq:ComputVarSolDoubleLimit}, the variance of the solution $u_{c=0}(t,x)$ in physical space of the heat equation will behave at long time, in the limit of vanishing small viscosities, as
\begin{align}\label{eq:VarSolSHEPhySpace}
\lim_{t\to\infty}\E \left[u_{c=0}^2(t,x)\right] \build{\sim}_{\nu\to 0}^{} \frac{1}{\nu}\int \frac{\widehat{C}_f(k)}{8\pi^2|k|^2}dk,
\end{align}
in any dimension $d$. Note that the right-hand side of \eqref{eq:VarSolSHEPhySpace} is positive and finite since the support of $\widehat{C}_f$ is assumed to be compact. Whereas it is expected that the kinetic energy of turbulent fluids is independent of viscosity, the one of the stochastic heat equation is inversely proportional to $\nu$. In other words, the stochastic heat equation is poorly efficient at dissipating energy compared to the Navier-Stokes equations and to our model (for $c>0$). This is due to the absence of the cascade phenomenon for the stochastic heat equation.

\section{Numerical method} \label{Sec:NumMethod}

This section is devoted to the description of the method employed for the numerical simulation of the model described above. That model can be studied in both physical \eqref{eq:DefEvolPhysSpace} and Fourier \eqref{eq:DefEvolFourSpace} formulations. We choose to discretize the formulation~\eqref{eq:DefEvolFourSpace} of the problem in the Fourier variables. This is motivated by the presence of the advection operator in the dynamics which plays a key role in the cascade phenomenon. In addition all the other terms in the dynamics can also be easily interpreted and computed in the formulation~\eqref{eq:DefEvolFourSpace}. Finally, discretizing the Fourier variable formulation is also natural since the boundary conditions~\eqref{eq:BCFourSpace} are imposed in that version of the problem.

It is worth mentioning that in this article we are mostly interested in the statistical behavior of the model, in the large time regime: it is therefore sufficient to propose numerical algorithms which are efficient for the approximation in distribution of the stochastic processes  (instead of their trajectories).

The numerical simulation of the stochastic evolution equation~\eqref{eq:DefEvolFourSpace}  requires to solve several non-trivial issues. Indeed, our objective is to study the long-time behavior of the system, in order to observe the power-law behavior of the generated fractional random fields. We thus need to ensure first that the numerical approximation reaches a stationary state, which depends on the chosen numerical scheme and on the numerical discretization parameters, and second that this stationary state is an accurate approximation of the stationary state of the system~\eqref{eq:DefEvolFourSpace}. Even for deterministic dynamical systems, crude methods may fail and the design of effective methods is a non trivial task. The addition of stochastic external forcing naturally introduces additional difficulties, see for instance \cite{CEB}. In this work, the fact that the external forcing is $\delta$-correlated in time $t$ and in the Fourier variable $k$ is an important challenge.

We first describe the discretization in the Fourier domain using the finite volume method, and we then describe the temporal discretization using a splitting method and exponential integrators for Ornstein--Uhlenbeck dynamics. Finally, we will describe the fully-discrete schemes. Note that in this article, we do not investigate the convergence properties of the scheme, this question is fundamental but is left for future works.

In the sequel, we study Fourier transforms $\hat{g}(k)$ of real-valued random fields $g(x)$. As a result, the Hermitian symmetry property $\overline{\hat{g}(k)}=\hat{g}(-k)$ is satisfied, where $\overline{\cdot}$ stands for the complex conjugate. The description of the correlations of $\hat{g}(k)$ requires in general to consider both $\E[\hat{g}(k)\hat{g}(k')]$ and $\E[\hat{g}(k)\overline{\hat{g}(k')}]$, for all $k,k'$. However owing to the Hermitian symmetry property giving the expression for $\E[\hat{g}(k)\overline{\hat{g}(k')}]$ for all $k,k'$ is sufficient.

\subsection{Spatial discretization: finite volume method}\label{Sec:FVmethod}

\subsubsection{General definition the finite volume method}

Since the external stochastic forcing $\hat{f}(t,k)$ is $\delta$-correlated in $(t,k)$, the solution $\hat{u}(t,k)$ of~\eqref{eq:DefEvolFourSpace} cannot be interpreted pointwise. However, for any bounded \textit{volume}, \textit{i.e.} any subset $\K\subset \mathcal{D}_{\kappa}=\{k\in\R^d\;~|k|\ge \kappa\}$ with volume $|\K|>0$ in the Fourier domain, one can average the forcing on $\K$ and set
\[
\hat{f}_\K(t)=\fint_{\K} \hat{f}= \frac{1}{|\K|} \int_\K \hat{f}(t,k) dk.
\]
With that definition, for each volume $\K$, it is straightforward to check that $\bigl(\hat{f}_{\K}(t)\bigr)_{t\ge 0}$ is a complex centered white noise process in time: %\ceb{pour le Dirac, $t_2-t_1$ entre parentheses ou en indice?}
\begin{align*}
\E[\hat{f}_{\K}(t_1) \overline{\hat{f}_{\K}(t_2)}]&= \delta(t_2-t_1) \frac{1}{|\K|^2} \int_{\K} \widehat{C}_f(k)dk,\\
\E[\hat{f}_{\K}(t_1) {\hat{f}_{\K}(t_2)}]&= \delta(t_2-t_1) \frac{1}{|\K|^2} \int_{\K\cap(-\K)} \widehat{C}_f(k)dk
\end{align*}
where $-\K=\{-k;~k\in\K\}$ is the symmetric of $\K$ with respect to the origin. Note that $f_{-\K}(t)=\overline{f_\K(t)}$.
%and we recall that, for convenience, $\E[\hat{f}_{\K}^2(t)]=0$,
More generally, if $\K_1$ and $\K_2$ are two volumes,  then, owing to~\eqref{eq:CorrForcingFourSpace}, one has
\begin{align*}
\E[\hat{f}_{\K_1}(t_1) \overline{\hat{f}_{\K_2}(t_2)}]&= \delta(t_2-t_1) \frac{1}{|\K_1 | | \K_2 |} \int_{\K_1\cap \K_2} \widehat{C}_f(k)dk,\\
\E[\hat{f}_{\K_1}(t_1) {\hat{f}_{\K_2}(t_2)}]&= \delta(t_2-t_1) \frac{1}{|\K_1 | | \K_2 |} \int_{\K_1\cap (-\K_2)} \widehat{C}_f(k)dk.
\end{align*}
Moreover, if $\K_1\cap\K_2=\emptyset$ and if $\K_1\cap(-\K_2)=\emptyset$, then $\bigl(\hat{f}_{\K_1}(t)\bigr)_{t\ge 0}$ and $\bigl(\hat{f}_{\K_2}(t)\bigr)_{t\ge 0}$ are independent complex white noise processes.

The observation above suggests to also average the solution over volumes: for any volume $\K$ consider
\begin{equation}\label{eq:DefGeneuK}
\hat{u}_\K(t)=  \fint_\K \hat{u}= \frac{1}{|\K|} \int_\K \hat{u}(t,k) dk,
\end{equation}
where $\hat{u}(t,k)$ denotes the solution of~\eqref{eq:DefEvolFourSpace}. The finite volume method consists in introducing a countable locally finite decomposition of the domain $\mathcal{D}_\kappa$ into volumes $\K$, and to propose an evolution equation for $\hat{u}_\K(t)$.

Let us first observe that integrating the evolution equation~\eqref{eq:DefEvolFourSpace} over an arbitrary volume $\K$ shows that $\hat{u}_\K(t)$ is the solution to the following equation:
\begin{equation}\label{eq:uK}
\begin{aligned}
\partial_t \hat{u}_\K &+ c\fint_\K \mbox{div}_{k} \left( \frac{k}{|k|} \hat{u} \right)
+   c\fint_\K \frac{( H + \frac{1}{2})}{|k|} \hat{u}\\
&=  - \nu\fint_\K (2\pi |k|)^2 \hat{u} + \hat{f}_\K.
\end{aligned}
\end{equation}
Using the Stokes formula, the second term of the left-hand side of~\eqref{eq:uK} is written as
\begin{align}\label{eq:StokesTransport}
\fint_\K  \mbox{div}_{k} \left( \frac{k}{|k|} \hat{u} \right) =  \frac{1}{|\K|}\int_{\partial \K} \frac{k \cdot {\bf n}}{|k |}  \hat{u}(t,k)dk,
\end{align}
where $\partial\K$ denotes the boundary of $\K$ and ${\bf n}$ denotes the outward unit normal vector at $k\in\partial\K$. 

\subsubsection{Finite volume meshes with radial symmetry}

Let us now choose the form and shape of the finite volume $\K$ which should be especially well adapted to the symmetry of our problem. In particular, we would like to simplify, without making any approximation, the average over $\K$ of the divergence term in the right-hand side of~\eqref{eq:StokesTransport}. To do so, it is convenient to use the spherical coordinate system. Recall that any $k\in\R^d$ such that $k\neq 0$ can be uniquely written as $k=|k|\theta$ with $\theta\in\mathbb{S}^{d-1}$, where $\mathbb{S}^{d-1}=\{k\in\R^d;~|k|=1\}$ is the unit sphere of dimension $d$, see below for some details in dimensions $d=1,2,3$.

\begin{figure*}
\begin{center}
\includegraphics[width=15cm]{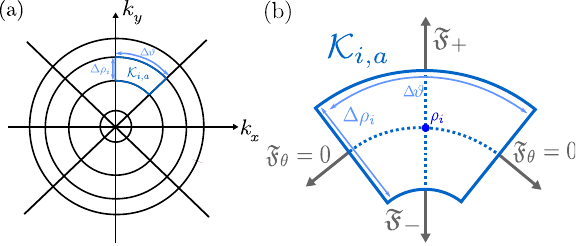}
\end{center}
\caption{\label{fig:Cartoon} (a) Representation of the discretization of the plane spanned by the wave vector $k$ in dimension $d=2$ using finite volumes method, each cell corresponds to a finite volume $\K_{i,a}$ (Eq. \ref{Meshelements}) with radial step $\Delta \rho_i$. (b) Representation of a unit cell $\K_{i,a}$ with our notation. We superimpose the non-vanishing radial fluxes (Eq. \ref{eq:RadialFluxes}). }
\end{figure*}

Using spherical coordinates, we now consider volumes $\K$ of the type
\begin{equation}\label{eq:SphericalSymFiniteVolume}
\K=\{k=|k|\theta;~\rho_{-}\le |k|\le \rho_{+},~\theta\in\Theta\},
\end{equation}
where $\rho_{-}\le \rho_{+}$ are two positive real numbers, and $\Theta$ is a subset of the unit sphere $\mathbb{S}^{d-1}$. For any volume $\K$ given by~\eqref{eq:SphericalSymFiniteVolume}, the boundary $\partial \K$ is decomposed into three parts
\begin{align*}
\partial_+ \K&=\{k=|k|\theta;~|k|=\rho_{+},~\theta\in\Theta\}\\
\partial_- \K&=\{k=|k|\theta;~|k|=\rho_{-},~\theta\in\Theta\}\\
\partial_\theta \K&=\{k=|k|\theta;~\rho_{-}<|k|<\rho_{+},~\theta\in\partial\Theta\}
\end{align*}
where $\partial\Theta$ denotes the boundary of the set $\Theta$. The boundary integral in the right-hand side of~\eqref{eq:StokesTransport} obtained by the application of the Stokes formula above can then be simplified:
\[
\frac{1}{|\K|}\int_{\partial \K} \frac{k \cdot {\bf n}}{|k |}  \hat{u}(t,k)dk=\mathfrak F_{+}+\mathfrak F_{-},
\]
where only the radial fluxes $\mathfrak F_{\pm}$
\begin{equation}\label{eq:RadialFluxes}
\mathfrak F_{\pm} = \frac{1}{|\K|}\int_{\partial_\pm \K} \frac{k \cdot {\bf n}}{|k|} \hat{u}(t,k)dk=\frac{\pm 1}{|\K|}\int_{\partial_\pm \K}\hat{u}(t,k)dk,
\end{equation}
accross $\partial_\pm\K$ contribute, whereas the flux $\mathfrak F_{\theta}$ across the transverse boundary $\partial_\theta\K$ vanishes, i.e.
\begin{equation}\label{eq:TransverseFluxes}
\mathfrak F_{\theta} =\int_{\partial_\theta \K} \frac{k \cdot {\bf n}}{|k |} \hat{u}(t,k)dk=0,
\end{equation}
since $k\cdot {\bf n}=0$ vanishes for $k\in\partial_\theta\K$. See Fig. \eqref{fig:Cartoon}(b) for an illustration. Therefore choosing the volumes $\K$ of the type considered above is natural: they are adapted to the radial symmetries of the operators and of the forcing. Moreover, this choice is adapted to describe the advection behavior in the radial variable $|k|$, which is exhibited in the identity \eqref{div-to-partial_k}.

It is worth mentioning that all the calculations performed so far are exact, in other words no approximation procedure has been introduced yet. To proceed further, let us describe more precisely the finite volume mesh that we will consider. Let $\bigl(\Theta_a\bigr)_{a\in\mathcal{A}}$ denote a finite volume decomposition of the sphere $\mathbb{S}^{d-1}$ where $\mathcal{A}$ is a finite set and assume that the volume of $|\Theta_a|$ does not depend on $a$. In addition, in order to preserve the Hermitian symmetry of the random fields in the $k$ variable the finite volume mesh also needs to satisfy some symmetry property. Instead of describing this in arbitrary dimension $d$, we give details about the cases $d=1,2,3$ below.

\subsubsection{Discretization of the unit sphere in dimensions $d=1,2,3$}\label{Par:DiscreteSd}

Let us make the description of the volumes $\Theta_a$ in the unit sphere more precise in dimension $d=1,2,3$.

In dimension $d=1$, any $k\neq 0$ can be written as $k=\text{sign}(k)|k|$, thus one can consider $\mathcal{A}=\{\pm 1\}$, and the volumes $\Theta_+=\{1\}$ and $\Theta_-=\{-1\}$ are symmetric with respect to $0$. In practice owing to the Hermitian symmetry property it is sufficient to deal with $\Theta_+$.

In dimension $d=2$, one has the polar decomposition $k=|k|(\cos{\vartheta},\sin{\vartheta})$ with angle $\vartheta\in[0,2\pi]$ for any $k\neq 0$. Let $\Delta\vartheta=\pi/N_\vartheta$ for some integer $N_\vartheta$, then one can consider $\mathcal{A}=\{0,\ldots,2N_{\vartheta}-1\}$, and for all $a\in\mathcal{A}$, one can set $\Theta_a=\{(\cos{\vartheta},\sin{\vartheta});~a\Delta\vartheta \le \vartheta \le (a+1)\Delta\vartheta\}$. Observe that one has $-(\cos(\vartheta),\sin(\vartheta))=(\cos(\vartheta+\pi),\sin(\vartheta+\pi))$, therefore for any $a\in\{0,\ldots,N_\vartheta-1\}$, the volumes $\Theta_{a+N_\vartheta}$ and $\Theta_a$ are symmetric with respect to $0$. In practice owing to the Hermitian symmetry property it is thus sufficient to deal with $\Theta_a$ for $a\in\{0,\ldots,N_\vartheta-1\}$.

In dimension $d=3$, one has the spherical decomposition $k=|k|(\cos{\vartheta},\sin{\vartheta}\cos{\varphi},\sin{\vartheta}\sin{\varphi})$ with angles $(\vartheta,\varphi)\in[0,\pi]\times[0,2\pi]$, for any $k\neq 0$. Note that one has $-(\cos{\vartheta},\sin{\vartheta}\cos{\varphi},\sin{\vartheta}\sin{\varphi})=(\cos(\pi-\vartheta),\sin(\pi-\vartheta)\cos(\varphi+\pi),\sin(\pi-\vartheta)\sin(\varphi+\pi))$. Given $\Delta\vartheta=\pi/(N_\vartheta)$ and $\Delta\varphi=\pi/N_\varphi$, with integers $N_\vartheta$ and $N_\varphi$, one can consider $\mathcal{A}=\{(a_\vartheta,a_\varphi);~0\le a_\vartheta \le N_\vartheta-1,~0\le a_\varphi\le 2N_\varphi-1\}$, and for all $a=(a_\vartheta,a_\varphi)\in\mathcal{A}$ one can set $\Theta_a=\{(\cos{\vartheta},\sin{\vartheta}\cos{\varphi},\sin{\vartheta}\sin{\varphi});~a_\vartheta\Delta\vartheta \le \vartheta \le (a_\vartheta+1)\Delta\vartheta,~a_\varphi\Delta\varphi \le \varphi \le (a_\varphi+1)\Delta\varphi\}$. The volumes $\Theta_{(N_\vartheta-1-a_\vartheta,a_\varphi+N_\varphi)}$ and $\Theta_a$ are symmetric with respect to $0$. In practice owing to the Hermitian symmetry property it is thus sufficient to deal with $\Theta_a$ for $a\in\{(a_\vartheta,a_\varphi);~0\le a_\vartheta\le N_\vartheta-1,~0\le a_\varphi\le N_\varphi-1\}$.

The description could be extended to higher dimensions, the details are omitted since we only deal with dimensions $d=1,2,3$ in practice.

\subsubsection{Discretization of the radial component}

It remains to make the discretization of the radial component $|k|$ of the wave vector $k$ more precise. Let $\bigl(\rho_{i-\frac12}\bigr)_{i\ge 1}$ be an increasing sequence of positive real numbers, and assume that $\rho_{\frac12}=\kappa$. Then for all $i\ge 1$ and all $a\in\mathcal{A}$ we set
\begin{equation}\label{Meshelements}
\K_{i,a}=\{k=|k|\theta;~\rho_{i-\frac12}\le |k|\le \rho_{i+\frac12},~\theta\in\Theta_a\}.
\end{equation}
See Fig. \eqref{fig:Cartoon}(a) for a representation of the discretization method using finite volumes $\K_{i,a}$ given by \eqref{Meshelements} in dimension $d=2$.

For all $i\ge 1$, define the radial step size
\begin{equation}
\Delta \rho_i =\rho_{i+\frac12}-\rho_{i-\frac12}.
\end{equation}
As will be explained below, it is crucial that $\Delta \rho_i$ depends on $i\ge 1$. Therefore the mesh is not uniform in the radial component, as can be observed in Fig. \eqref{fig:Cartoon}(a). The values of $\rho_{i+\frac12}$ and of $\Delta \rho_i$ are imposed below depending on the time step size $\Delta t$.
In Fig. \eqref{fig:Cartoon}(b), we represent a volume $\K_{i,a}$ \eqref{Meshelements} with the corresponding non-vanishing radial fluxes \eqref{eq:RadialFluxes} and the vanishing transverse fluxes \eqref{eq:TransverseFluxes}.

\subsubsection{Approximation of the integrals}

To define the finite volume semi-discrete scheme, it is necessary to approximate the integrals appearing in~\eqref{eq:uK} for each volume $\K=\K_{i,a}$. First, we employ the upwind scheme \cite{CouIsa52} to deal with the advection (recall that $c>0$): we obtain the following approximations for the radial fluxes \eqref{eq:RadialFluxes}, for $i\ge 1$ and $a\in\mathcal{A}$, 
\begin{align*}
 \frac{1}{|\K_{i,a}|} \int_{\partial_+ \K_{i,a}}  \hat{u}(t,k)dk &\approx   \frac{| \partial_+ \K_{i,a}|}{|\K_{i,a}|} \hat{u}_{i,a}(t),\\
 \frac{1}{|\K_{i,a}|} \int_{\partial_- \K_{i,a}}  \hat{u}(t,k)dk &\approx   \frac{| \partial_- \K_{i,a}|}{|\K_{i,a}|} \hat{u}_{i-1,a}(t).
\end{align*}
Note that the volumes of $\K_{i,a}$ and $\partial_{\pm}\K_{i,a}$ are given by
\begin{align*}
|\K_{i,a}|&=|\Theta_a|\frac{\rho_{i+\frac12}^{d}-\rho_{i-\frac12}^{d}}{d},\\
|\partial_{\pm} \K_{i,a}|&=|\Theta_a|\rho_{i\pm\frac12}^{d-1}.
\end{align*}
As a result, one obtains the following approximation for the advection term~\eqref{eq:StokesTransport} that enters in the evolution~\eqref{eq:uK}: 
\begin{align*}
c\fint_{\K_{i,a}} \mbox{div}_k \left( \frac{k}{|k|} \hat{u} \right) &\approx c\frac{| \partial_+ \K_{i,a}|}{|\K_{i,a}|} \hat{u}_{i,a}-c\frac{| \partial_- \K_{i,a}|}{|\K_{i,a}|} \hat{u}_{i-1,a}
 \\
&=c\frac{d\rho_{i-\frac12}^{d-1}}{\rho_{i+\frac12}^{d}-\rho_{i-\frac12}^{d}}\bigl(\hat{u}_{i,a}-\hat{u}_{i-1,a}\bigr)\\
&\quad+cd\frac{\rho_{i+\frac12}^{d-1}-\rho_{i-\frac12}^{d-1}}{\rho_{i+\frac12}^{d}-\rho_{i-\frac12}^{d}}\hat{u}_{i,a}.
\end{align*}
The last expression above can be written as
\begin{equation}\label{eq:ApproxAdvection}
c\frac{\hat{u}_{i,a}-\hat{u}_{i-1,a}}{h_i} + d_i \hat{u}_{i,a}
\end{equation}
with auxiliary parameters $h_i>0$ and $d_i>0$ defined for all $i\ge 1$ by
\begin{equation}\label{eq:hi-di}
\begin{aligned}
h_i&=\frac{\rho_{i+\frac12}^{d}-\rho_{i-\frac12}^{d}}{\rho_{i-\frac12}^{d-1}},\\
d_i&=cd\frac{\rho_{i+\frac12}^{d-1}-\rho_{i-\frac12}^{d-1}}{\rho_{i+\frac12}^{d}-\rho_{i-\frac12}^{d}}.
\end{aligned}
\end{equation}
In the approximation~\eqref{eq:ApproxAdvection} of the advection term from~\eqref{eq:uK}, the first term $c\frac{\hat{u}_{i,a}-\hat{u}_{i-1,a}}{h_i}$ accounts for the advection in the radial coordinate whereas the second term $d_i \hat{u}_{i,a}$ can be interpreted as a numerical dissipation term ($d_i\ge 0$) due to the numerical approximation procedure. This is not a physical dissipation but appears in the numerical method due to considering curved mesh elements $\K_{i,a}$.
Note that the values of $h_i$ and $d_i$ depend on the choice of the sequence $\bigl(\rho_{i-\frac12}\bigr)_{i\ge 1}$.

For the other integral terms appearing in~\eqref{eq:uK}, a midpoint approximation is applied: for all $i\ge 1$, set
\begin{align}\label{eq:midpointapproximation}
\rho_i=\frac{\rho_{i-\frac12}+\rho_{i+\frac12}}{2},
\end{align}
then we consider the approximations 
\begin{align*}
\fint_{\K_{i,a}} \frac{( H + \frac{1}{2})}{|k|} \hat{u}&\approx \frac{( H + \frac{1}{2})}{\rho_i} \hat{u}_{i,a}\\
\fint_{\K_{i,a}}|k|^2\hat{u}&\approx \rho_i^2\hat{u}_{i,a}.
\end{align*}

\subsubsection{Final spatial discretization method}

Finally, combining the approximations above we obtain the finite volume method: for $t\ge 0$, $i\ge 1$ and $a\in\mathcal{A}$, one has
\begin{equation}\label{eq:FVsemi}
\partial_t \hat{u}_{i,a}(t) + c\frac{\hat{u}_{i,a}(t)-\hat{u}_{i-1,a}(t)}{h_i} + D_i \hat{u}_{i,a}(t) = \hat{f}_{i,a}(t),
\end{equation}
where 
$$\hat{f}_{i,a}(t)=\hat{f}_{\K_{i,a}}(t)=  \fint_{\K_{i,a}}\hat{f}(t,k)dk$$
 and with the auxiliary parameters $D_i$ defined for all $i\ge 1$ by
\[
D_i = d_i +\frac{( H + \frac{1}{2})}{\rho_i}+ \nu (2 \pi \rho_i)^2.
\]
The expression $D_i \hat{u}_{i,a}(t)$ in~\eqref{eq:FVsemi} can be interpreted as a dissipation term, which combines several effects. Recall that $d_i$ is not a physical dissipation term, it results from the numerical approximation procedure. On the contrary, the two other terms in the definition of $D_i$ have a physical meaning as dissipation terms.

The finite volume method~\eqref{eq:FVsemi} needs to be supplemented with initial and boundary conditions
\begin{align*}
&\hat{u}_{i,a}(0)=0,~i\ge 1,~a\in\mathcal{A},\\
&\hat{u}_{0,a}(t)=0,~t\ge 0,~a\in\mathcal{A}.
\end{align*}
The structure of the stochastic forcing $\bigl(\hat{f}_{i,a}\bigr)_{i\ge 1,a\in\mathcal{A}}$ is simple: these are independent $\delta$-correlated temporal white noise processes: one has
\begin{equation}\label{corr-fia}
\E[\hat{f}_{i,a}(t) \overline{\hat{f}_{j,b}(s)}]=\delta(t-s)\delta_{i,j}\delta_{a,b} \frac{1}{|K_{i,a}|^2} \int_{K_{i,a}} \widehat{C}_f(k)dk,
\end{equation}
where we have used the notation $\delta_{i,j}$ for the Kronecker delta function. In the finite volume method~\eqref{eq:FVsemi}, the advection behavior only takes place in the $|k|$ variable in~\eqref{eq:DefEvolFourSpace} and this is represented in the numerical method~\eqref{eq:FVsemi} by the index $i$, whereas $a\in\mathcal{A}$ can be considered as a parameter.

So far, we have only taken into account the spatial discretization in the construction of the finite volume method~\eqref{eq:FVsemi}. It remains to deal with the temporal discretization.

\subsection{Temporal discretization: splitting integrator}

Let us denote by $\Delta t>0$ the time step. The objective is to define a computable approximation denoted by $\hat{u}_{i,a}^n$ of $\hat{u}_{i,a}(t_n)$, at discrete times $t_n=n\Delta t$ for integers $n\ge 0$. In order to propose an approximation scheme for~\eqref{eq:FVsemi} which is able to capture accurately the stationary state of the exact solution, we cannot rely on the standard explicit Euler scheme. We propose to apply a splitting method, see for instance~\cite{blanes,MR2009376}: this consists in decomposing the evolution into subsystems which can be solved exactly or approximately (starting from any initial condition) and in combining the results at each time step to define a numerical approximation of the full system.

The evolution equation for the finite volume scheme~\eqref{eq:FVsemi} contains three terms: an advection term, a dissipation term, and a stochastic forcing term. There are several possible choices to combine them. The choice we propose in this article is made to treat carefully all the three contributions.

On the one hand, we consider the dissipation and stochastic forcing terms together (and the advection term is omitted): we obtain the subsystems
\begin{equation}\label{eq:splitOU}
\partial_t \hat{u}^{sd}_{i,a}(t) + D_i \hat{u}^{sd}_{i,a}(t) = \hat{f}_{i,a}(t),
\end{equation}
parameterized by $i\ge 1$ and $a\in\mathcal{A}$ (with boundary condition $\hat{u}^{sd}_{0,a}(t)=0$ for all $t\ge 0$ and $a\in\mathcal{A}$). Observe that~\eqref{eq:splitOU} is a system of independent Ornstein--Uhlenbeck dynamics, independence being considered up to the Hermitian symmetry property. For the subsystem~\eqref{eq:splitOU} we propose an exponential integrator in time which is exact in distribution.

On the other hand, we consider the advection term only (and the dissipation and stochastic forcing terms are omitted): we obtain the subsystem
\begin{equation}\label{eq:splitAdv}
\partial_t \hat{u}^{ad}_{i,a}(t) + c\frac{\hat{u}^{ad}_{i,a}(t)-\hat{u}^{ad}_{i-1,a}(t)}{h_i} =0,
\end{equation}
for $i\ge 1$, $a\in\mathcal{A}$, with the boundary condition $\hat{u}^{ad}_{0,a}(t)=0$. Note that~\eqref{eq:splitAdv} is a deterministic system. For the system~\eqref{eq:splitAdv} we propose an explicit Euler scheme for temporal discretization, which turns out to represent exactly the advection phenomenon in the fully-discrete setting.

Below we explain the resolution of the two subsystems~\eqref{eq:splitOU} and~\eqref{eq:splitAdv} and then how to combine the solutions in a splitting scheme.

\subsubsection{Integration of the Ornstein--Uhlenbeck dynamics~\eqref{eq:splitOU}}\label{Sec:SplitOU}

Given the solution $\hat{u}^{sd}(t_n)$ at time $t_n$, the solution of~\eqref{eq:splitOU} at time $t_{n+1}=t_n+\Delta t$ has the following expression:
\[
\hat{u}^{sd}_{i,a}(t_{n+1})=e^{-\Delta tD_i}\hat{u}^{sd}_{i,a}(t_n)+\int_{t_n}^{t_{n+1}}e^{-(t_{n+1}-s)D_i}\hat{f}_{i,a}(s)ds.
\]
To define an algorithm which allows to compute the value of $\hat{u}^{sd}_{i,a}(t_{n+1})$ as a function of $\hat{u}^{sd}_{i,a}(t_n)$, it is sufficient to give the distribution of the random variables
\[
\hat{g}_{i,a}^{n}=\int_{t_n}^{t_{n+1}}e^{-(t_{n+1}-s)D_i}\hat{f}_{i,a}(s)ds.
\]
Indeed, as previously mentioned we are interested only in the statistical properties of the spatio-temporal random fields. Being able to sample $\hat{u}^{sd}_{i,a}(t_{n+1})$ from $\hat{u}^{sd}_{i,a}(t_n)$ and $\hat{g}_{i,a}^{n}$ is a consequence of the Markov property for the Ornstein--Uhlenbeck dynamics. Since the forcing is $\delta$-correlated in time and in the variables $i$ and $a$, see \eqref{corr-fia}, the $\hat{g}_{i,a}^{n}$ are independent centered Gaussian random variables: using~\eqref{corr-fia} the correlation structure is given by
\[
\E\bigl[\hat{g}_{i,a}^{n} \overline{\hat{g}_{j,b}^{m}}\bigr]=\delta_{m,n}\delta_{i,j}\delta_{a,b} \frac{1- e^{-2 \Delta t D_i}}{2 D_i} \frac{1}{|\K_{i,a}|^2} \int_{\K_{i,a}} \widehat{C}_f.
\]
Since the volume $|\K_{i,a}|$ of $\K_{i,a}$ only depends on $i$ and does not depend on $a$, and since $\widehat{C}_f$ only depends on the radial component $|k|$ of $k$, one has
\[
\E\bigl[\hat{g}_{i,a}^{n} \overline{\hat{g}_{j,b}^{m}}\bigr]=\delta_{m,n}\delta_{i,j}\delta_{a,b} \varrho_i^2
\]
where
\[
\varrho_i=\sqrt{\frac{1-e^{-2\Delta t D_i}}{2D_i|\K_{i,a}|^2} \int_{K_{i,a}} \widehat{C}_f}.
\]
In other words, one has the equality in distribution
\[
\hat{g}_{i,a}^{n}=\varrho_i\hat{\gamma}_{i,a}^{n}
\]
where $\bigl(\hat{\gamma}_{i,a}^{n}\bigr)_{i\ge 1,a\in\mathcal{A},n\ge 0}$ are independent standard complex Gaussian random variables (independence being understood up to Hermitian symmetry):
\[
\E[\hat{\gamma}_{i,a}^{n}\overline{\hat{\gamma}_{j,b}^{m}}]=\delta_{m,n}\delta_{i,j}\delta_{a,b}.
\]
Setting $\hat{u}^{n,sd}_{i,a}=0$ and for all $n\ge 0$
\begin{equation}\label{eq:splitOU-solution}
\hat{u}^{n+1,sd}_{i,a}=e^{-\Delta tD_i}\hat{u}^{n,sd}_{i,a}+ \varrho_i \hat{\gamma}_{i,a}^{n},
\end{equation}
provides an algorithm which computes exactly in distribution the solution of the Ornstein--Uhlenbeck subsystem~\eqref{eq:splitOU}: one has the equality in distribution
\[
\hat{u}^{n,sd}_{i,a} \underset{\text{(law)}}{=}\hat{u}^{sd}_{i,a}(n\Delta t),
\]
or all $i\ge 1$, $a\in\mathcal{A}$ and $n\ge 0$ and for any choice of the time-step size $\Delta t$.

Using the exact simulation in distribution~\eqref{eq:splitOU-solution} of the Ornstein--Uhlenbeck dynamics~\eqref{eq:splitOU} is elementary and allows us to capture correctly the asymptotic behavior of the model. This would fail if for instance the explicit or the implicit Euler schemes were considered. We refer for instance to \cite{CEB} for a description of this issue when discretizing the stochastic heat equation.

\subsubsection{Integration of the discrete advection dynamics~\eqref{eq:splitAdv}}\label{Chap:DiscretAdvecDyn}

The deterministic system of equations~\eqref{eq:splitAdv} can be approximated using the explicit Euler scheme: for all $n\ge 0$, one obtains
\begin{equation}\label{eq:DiscretAdvectDyn}
\hat{u}_{i,a}^{n+1,ad}-\hat{u}_{i,a}^{n,ad}+\frac{c\Delta t}{h_i}\bigl(\hat{u}_{i,a}^{n,ad}-\hat{u}_{i-1,a}^{n,ad}\bigr)=0,
\end{equation}
for all $i\ge 1$ and $a\in\mathcal{A}$. The fully-discrete scheme~\eqref{eq:DiscretAdvectDyn} can be interpreted as the upwind scheme applied to the auxiliary radial advection equation of a given field $\hat{v}^{ad}(t,|k|,\theta)$ which would read
\begin{equation}\label{eq:advectionauxiliary}
\partial_t\hat{v}^{ad}(t,|k|,\theta)+c\partial_{|k|}\hat{v}^{ad}(t,|k|,\theta)=0
\end{equation}
parametrized by $\theta$, with time-step size $\Delta t$ and with a  mesh size $h_i$ which will be chosen to be non constant.

Usually, the stability of the scheme~\eqref{eq:DiscretAdvectDyn} above is ensured when the Courant--Friedrichs--Lewy (CFL) condition
\[
c \frac{\Delta t}{h_i}\le 1,~\forall~i\ge 1,
\]
is satisfied. In addition, numerical dissipation occurs if the equality does not hold. In order to capture the long-time behavior and the low regularity properties of the random field, it is desirable to minimize numerical dissipation. This leads us to impose the condition
\begin{equation}\label{eq:CFL1}
c \frac{\Delta t}{h_i}=1,
\end{equation}
for all $i\ge 1$.
Observe that when the condition~\eqref{eq:CFL1} is satisfied, then the expression of the solution of the numerical scheme~\eqref{eq:DiscretAdvectDyn} is simple: one has for all $n\ge 0$ and all $i\ge 1,a\in\mathcal{A}$
\begin{equation}\label{eq:splitAdv-solution}
\hat{u}_{i,a}^{n+1,ad}=\hat{u}_{i-1,a}^{n,ad},
\end{equation}
taking into account the boundary conditions $\hat{u}_{0,a}^{n,ad}=0$. The algorithm to compute $\hat{u}_{i,a}^{n+1,ad}$ from $\hat{u}_{i,a}^{n,ad}$ is straightforward, and one can compute all values of $\hat{u}_{i,a}^{n,ad}$ once the initial values $\hat{u}_{i,a}^{0,ad}$ are imposed.

Notice that the explicit Euler scheme~\eqref{eq:DiscretAdvectDyn} does not provide the exact solution of the advection subsystem~\eqref{eq:splitAdv}, even if the condition~\eqref{eq:CFL1} is satisfied. However, in fact it provides the exact solution of the auxiliary advection problem~\eqref{eq:advectionauxiliary} at grid points of the spatio-temporal mesh.

The condition~\eqref{eq:CFL1} implies that $h_i=c\Delta t$ does not depend on $i$. Using the definition~\eqref{eq:hi-di} of $h_i$, one obtains the recursion formula for $\rho_{i+\frac12}$
\begin{equation}\label{mesh-generation}
\rho_{i+\frac12}^d=\rho_{i-\frac12}^{d}+h_i\rho_{i-\frac12}^{d-1}=\rho_{i-\frac12}^{d}\bigl(1+\frac{c\Delta t}{\rho_{i-\frac12}}\bigr),
\end{equation}
for all $i\ge 1$, where $\rho_{\frac12}=\kappa$. Therefore the mesh is not uniform if $d\ge 2$, \textit{i.e.} $\Delta\rho_i$ depends on $i\ge 1$. Choosing a constant $h_i$ for numerical reasons thus imposes the geometry of the mesh, which is not standard in the scientific computing literature.

\subsubsection{Splitting integrator}

We have now presented all the ingredients in order to provide the definition of the fully-discrete scheme. We consider the mesh given by~\eqref{Meshelements} and~\eqref{mesh-generation}, and we recall that the time-step size is denoted by $\Delta t$ and that $t_n=n\Delta t$. Assume that the condition~\eqref{eq:CFL1} is satisfied. The initial value is given by
\[
\hat{u}_{i,a}^{0}=0,\quad \forall~i\ge 1,a\in\mathcal{A},
\]
but one may also consider more general initial values. In addition boundary conditions
\[
\hat{u}_{0,a}^{n}=0
\]
are also imposed for all $n\ge 1$.

We consider a Lie--Trotter splitting scheme, which combines two steps. Assume that the approximation $\bigl(\hat{u}_{i,a}^{n}\bigr)_{i\ge 1,a\in\mathcal{A}}$ has been computed. Then $\hat{u}_{i,a}^{n+1}$ is given by
\begin{equation}\label{eq:scheme}
\left\lbrace
\begin{aligned}
\hat{u}_{i,a}^{n+\frac12}&=e^{-\Delta tD_i}\hat{u}^{n}_{i,a}+ \varrho_i \, \hat{\gamma}_{i,a}^{n}\\
\hat{u}_{i,a}^{n+1}&=\hat{u}_{i-1,a}^{n+\frac12},
\end{aligned}
\right.
\end{equation}
for all $i\ge 1$ and $a\in\mathcal{A}$, using~\eqref{eq:splitOU-solution} in the first step and~\eqref{eq:splitAdv-solution} in the second step.

Note that at each iteration only linear operations on the solutions and addition of independent Gaussian random variables are performed. As a result, the numerical scheme is a Markov and Gaussian process.

We recall that in this article we are not interested in proving rigorous convergence results when the time-step and the mesh sizes vanish. We investigate in the next section the behavior of the scheme and we show that it is able to capture the power-law behavior predicted by theoretical analysis for the power spectral density \eqref{eq:ComputPSDDoubleLimit} and for the second order structure function \eqref{eq:ComputSecondOrderSFModelDoubleLimitSmallScales}), as well as the asymptotic behavior of the velocity variance \eqref{eq:ComputVarSolDoubleLimit}.

\section{Presentation of the numerical results}

\subsection{General comments regarding simulation in dimensions $d=1,2$ and $3$}\label{Par:GenComNS}

In the sequel, recalling previous developments presented in Section \ref{Sec:NumMethod}, we will be conducting numerical simulations of the evolution depicted in Eq. \ref{eq:uK}, for a finite viscosity $\nu>0$, where the wave vectors $k$ are discretized according to the finite volume method in the spherical symmetry pointed in Eq. \ref{eq:SphericalSymFiniteVolume}. In all subsequent simulations, we make the choice to take the mesh size $h_i=h$ (Eq. \ref{eq:hi-di}) along the radial direction to be constant, i.e. independent of the index $i$, and will be expressed in units of the cut-off parameter $\kappa$. Also, we will be working with a unit Courant number (Eq. \ref{eq:CFL1}), such that the discrete advection dynamics (Eq. \ref{eq:DiscretAdvectDyn}) is solved in an exact fashion according to the relation pointed in Eq. \ref{eq:splitAdv-solution}. As a consequence, the time step $\Delta t$ is automatically set to the value $h/c$ following Eq. \ref{eq:CFL1}. The remaining linear part of the dynamics, within the splitting approach described in Section \ref{Sec:SplitOU}, will be eventually solved according to the exponential scheme defined in Eq. \ref{eq:splitOU-solution}. The full integration procedure is summarized by the two steps described in \eqref{eq:scheme}. Hereafter, without loss of generality, all simulations will be conducted with the value $c=1$. Henceforth, to simplify the notation, we will be using often the misuse of language $\hat{u}_{\K}(t)$ to refer to the fully discretized version of the finite-volume Fourier mode $\hat{u}_{i,a}^{n}$ at time $t=t_n$ for the cell $\K=\K_{i,a}$ located at the radial coordinate $\rho_i$ and the angular coordinate $a$.

Notice that choosing the mesh size $h_i=h$ (Eq. \ref{eq:hi-di}) to be constant, in units of the cut-off parameter $\kappa$, is different from assuming a constant radial step size $\Delta \rho_i = \rho_{i+\frac12}-\rho_{i-\frac12}$, as it is  represented in Fig. \eqref{fig:Cartoon}(a). It is nonetheless true that the correspondance $h=\Delta \rho_i=\Delta \rho$ is exact in dimension $d=1$, but it is no more the case for $d\ge 2$. Thus, to compute $\rho_i$ under the midpoint approximation \eqref{eq:midpointapproximation}, we use the recursion relation \eqref{mesh-generation} to compute $\rho_{i-\frac12}$ and $\rho_{i+\frac12}$, which yields $\Delta \rho_i$.

Furthermore, the discretization procedure of the unit sphere $\mathbb{S}^{d-1}$ is detailed in Paragraph \ref{Par:DiscreteSd}, and finally, we enforce the Hermitian symmetry of $\hat{u}_{\K}(t)$ at each time step while time evolving the relevant Fourier modes, and completing if necessary the remaining modes located at the opposite side of the origin.

\subsection{Estimation of the relevant statistical quantities taking time and volume averaging}\label{Par:AveragProc}

We will first estimate the power spectral density $E(t,k)$ defined in \eqref{eq:DefPSD}, where the expectation is taken over the instances of the random forcing. This is crucial for the statistical characterization of the regularity of the solution, and it consists of several steps.

First of all, for a given value of the viscosity $\nu$, the solution reaches a statistically stationary state as proved in Ref. \cite{ApoBec23}, in which the variance of the solution becomes finite, and independent of viscosity for sufficiently low viscosities (Eq. \ref{eq:ComputVarSolDoubleLimit}). Starting from a vanishing initial condition, the characteristic time scale at which the system reaches this statistically stationary state should exclusively depend of $\nu$ if all wave numbers were numerically accessible. For a realistic simulation with a finite number of accessible wave numbers, viscosity has to be chosen such that spectral energy of the highest accessible wave number is negligible, following an exponential decrease as expected from the formal solution written in Eq. \ref{eq:MildSolution}. It is easy to get convinced that the underlying linear transport mechanism that enters in the present dynamics  (Eq. \ref{eq:DefEvolFourSpace}) would transfer an initially injected amount of energy at low wave numbers towards high wave numbers $|k|$ in a time of order $|k|/c$, such that, at the very best, the system will take a characteristic time of order $T^\star=k_{\text{max}}/c$ to reach the end of the spectral domain, where $k_{\text{max}}=\rho_N$, with $N$ being the number of finite volumes $\K$ along the radial direction. Chosen values of viscosity will be such that energy at $k_{\text{max}}$ is negligible, i.e. exponentially small, compared to say energy at low wave numbers, thus $T^\star$ can be considered as an optimistic upper-bound for the beginning of the statistically stationary state. 

Hereafter, estimation of various expectations entering in forthcoming statistical quantities will be based on empirical averages starting at time $T^\star$, each instances in these empirical averages will be taken at various instant such that they can be considered, in good approximation, as being statistically independent. The time lag between these samples will be specified later when discussing our numerical results in various space dimensions.

Because we only have access to the finite volume averaged $\hat{u}_{\K}(t)$ Fourier mode \eqref{eq:DefGeneuK} of the velocity field, we need to specify its relationship with the genuine PSD $E(t,k)$ defined in \eqref{eq:DefPSD}. We thus define the periodogram $E_\K (t)$ as the expectation of the square of the amplitude of $\hat{u}_{\K}(t)$, which is linked to the PSD $E(t,k)$ as
\begin{align}\label{eq:DefEKPSD}
E_\K (t) &\equiv \E \left| \hat{u}_{\K}(t)\right|^2\notag\\
&=  \frac{1}{|\K|^2} \int_\K E(t,k) dk,
\end{align}
that follows from the expressions given in \eqref{eq:DefPSD} and \eqref{eq:DefGeneuK}. As a consequence of \eqref{eq:DefEKPSD}, using the exact expression of $E(t,k) $ obtained in \eqref{eq:ComputPSDDoubleLimit} which eventually behaves proportionally to a pure power-law $|k|^{-(2H+d)}$ as $|k|\to \infty$ in the statistically stationary state and when $\nu\to 0$, $|\K| E_\K$ is also expected to behave as a similar power-law $\rho_i^{-(2H+d)}$ if the finite volume $\K$ is far from the origin, i.e. when $\rho_i\to\infty$, with a remaining multiplicative factor that can be derived from \eqref{eq:DefEKPSD}. In the sequel, because of the statistical isotropy in Fourier space, we will only display angle-averaged versions $E^{\Theta}_\K$ of $E_\K$ defined by 
\begin{align}\label{eq:DefEKPSDTheta}
E^{\Theta}_\K (t) =  \frac{1}{\Omega_d}\int_{\Theta\in\mathbb S^{d-1}} E_{\K}(t),
\end{align}
where $\Omega_d$ is the surface of the unit sphere in dimension $d$, given explicitly by the formula
\begin{align}\label{eq:SurfUnitSphereDimD}
\Omega_d=\frac{2\pi^{d/2}}{\Gamma(d/2)},
\end{align}
with $\Gamma$ standing for the usual gamma function.

\subsection{Back to physical space and averaging procedure for the second-order structure function}

Of tremendous importance from the physical point of view is the evaluation of the velocity field  in the physical space as the inverse Fourier transform of the finite volume Fourier modes $\hat{u}_{\K}(t)$. Recall first that, as a function of a continuous wave vector, the finite-volume Fourier mode $\hat{u}_{\K}(t)$ is a piecewise-constant function of the coarse-grained distributional Fourier transform $\hat{u}(t,k)$ \eqref{eq:DefGeneuK}. Furthermore, in space-dimension $d\ge 2$, because the shape and volume of the finite volume $\K$ depend on its location in the spectral domain, see for instance the cartoon displayed in  Fig. \eqref{fig:Cartoon}, we cannot define the coarse-graining procedure over $\K$ as a convolution of $\hat{u}(t,k)$ with a given windowing function. For these reasons, the relationship between the inverse Fourier transform of the piecewise-constant function $\hat{u}_{\K}(t)$ and the continuous field $u(t,x)$ is not obvious.

Nonetheless, going back to the discrete formulation, we propose to define the following field $\widetilde{u}_{\Delta}(t,x) $, at a given time $t$ and for a given position $x\in\R^d$,
\begin{align}\label{eq:DeftildeUDeltaPhysSpace}
\widetilde{u}_{\Delta}(t,x) = \sum_{n=1}^N\sum_{a\in\mathcal A}  e^{2i\pi x\cdot k_n }\hat{u}_{\K_{n,a}}(t)\rho_n^{d-1}\Delta \rho_n\Delta \Theta_a,
\end{align}
where $\mathcal A$ is the discretized subset of $\mathbb S^{d-1}$, defined in paragraph \ref{Par:DiscreteSd}, $k_n=\rho_n\theta$ with $\theta\in\Theta$, the radial  resolution $\Delta \rho_n$, and the corresponding differential solid angle $\Delta \Theta_a$ at the angular coordinate $a$. The validity of that approximation is a non-trivial question that we do not treat in this work. Note that in dimension $d\ge 2$, $\rho_n^{d-1}$ goes to infinity when $n$ increases due to imposing the condition \eqref{eq:CFL1} on $h_i$, therefore the convergence when $N\to\infty$ in \eqref{eq:DeftildeUDeltaPhysSpace} needs to be considered with care.

Then, following a similar time averaging procedure as for the estimation of the PSD that is described in the previous paragraph, we define the respective second-order structure function as follows:
\begin{align}\label{eq:DefS2tildeUDelta}
\left\langle \left(\delta_\ell \widetilde{u}_{\Delta} \right)^2\right\rangle \equiv \left\langle \left(\widetilde{u}_{\Delta}(t,x+\ell)-\widetilde{u}_{\Delta}(t,x)\right)^2\right\rangle,
\end{align}
where the brackets $\langle \cdot \rangle$ stand for the empirical average of the expectation over time, as previously done for the periodograms, but also over space, \emph{i.e.} averaging over all positions $x$ at which the field $\widetilde{u}_{\Delta}$ \eqref{eq:DeftildeUDeltaPhysSpace} is computed. As we will see, positions $x$ will be distributed over a Cartesian grid, using a uniform resolution $\Delta x$, that will be chosen in units of $k_{\text{max}}^{-1}$ in every direction, where $k_{\text{max}}$ is the largest accessible wave number.

\begin{figure}
\begin{center}
\includegraphics[width=8.5cm]{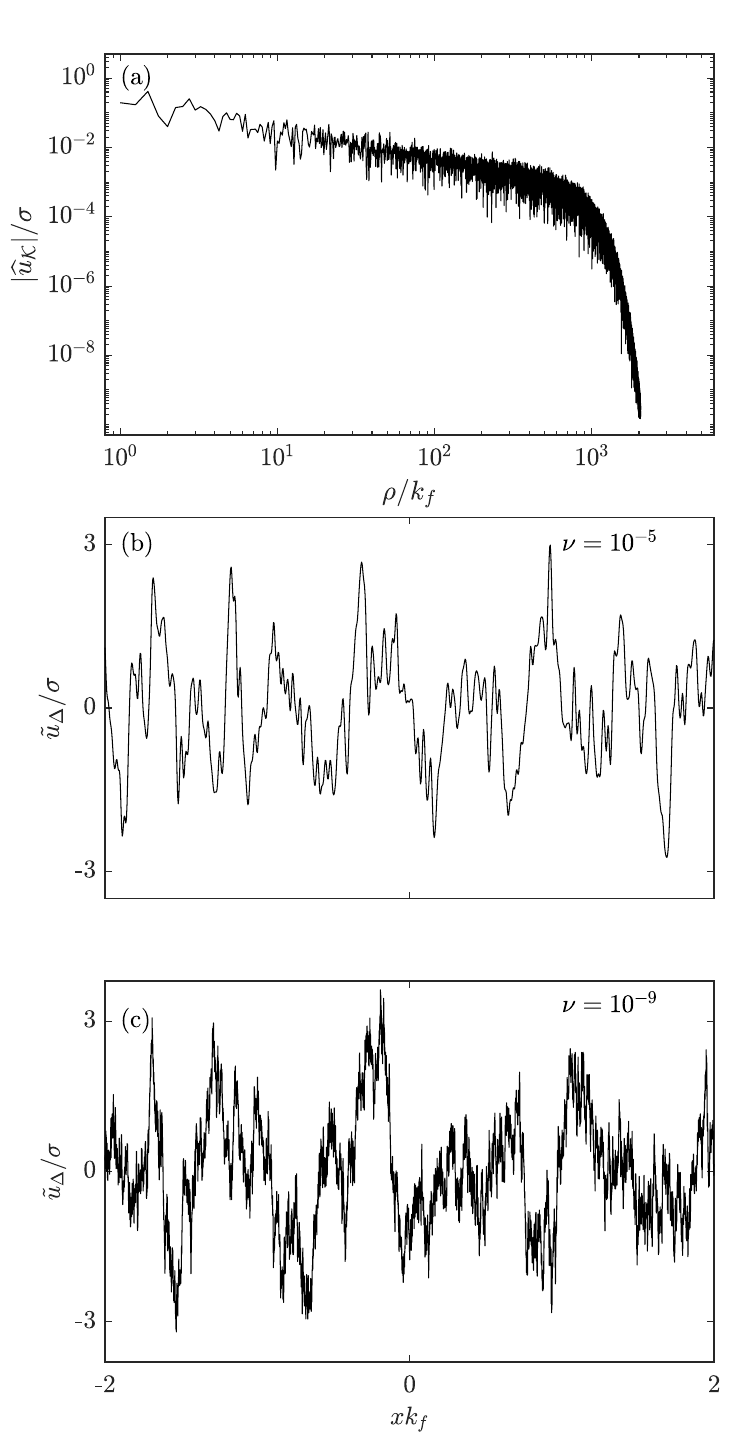}
\end{center}
\caption{\label{fig:Phys1D} Solution to the dynamics in Fourier space and physical space in the statistically stationary regime. All simulations are conducted with $H=1/3$, $c=1$, $\Delta \rho = h=\kappa=2^{-3}$ and $\widehat{C}_f(k)= \mathds{1}_{\kappa\leq |k| \leqslant k_f}$ (see \eqref{corr-fia}), with $k_f=4\kappa$. (a) Volume-averaged Fourier mode amplitude $|\hat{u}_{\K_{n,a}}(t)|$ in the statistically stationary regime (i.e. for $t>T^\star$), choosing $\nu=10^{-9}$ and using $N= 2^{12}$ collocation points in the radial direction, as a function of the radial coordinate $\rho_i$ \eqref{eq:midpointapproximation}. (b), (c) Physical space representation of the solution in the statistically stationary regime for $\nu=10^{-5}$ and $10^{-9}$ using correspondingly  $N= 2^7$ and $2^{12}$, at a given time in the statistically stationary regime, as a function of the non dimensional variable $xk_f$. The physical space representations are obtained using the inversion formula   \eqref{eq:DeftildeUDeltaPhysSpace1D} over $|x| \leqslant L_{\text{tot}}/2$, with the spatial resolution $\Delta x= 1/k_{\text{max}}$ with $k_{\text{max}}=\kappa+N\Delta \rho$, and the total length $L_{\text{tot}}$ of the physical domain chosen to be $L_{\text{tot}}=1/\Delta \rho$. }
\end{figure}

\subsection{Experiments and comments}

\subsubsection{One dimensional ($d=1$) simulations}

We conduct numerical simulations as described in paragraph \ref{Par:GenComNS}, and we begin performing these simulations in space dimension $d=1$. In Fig. \eqref{fig:Phys1D}(a), we display the absolute value of a  snapshot of the volume-averaged spectral field $\hat{u}_{\K_{n,a}}(t)$ in the statistically stationary regime which is reached after a transient at time $t>T^\star$, where $T^\star$ is defined and commented in paragraph \ref{Par:AveragProc}. Relevant additional parameters of the simulations are provided in the caption of Fig. \eqref{fig:Phys1D}.

As we can see in Fig. \eqref{fig:Phys1D}(a), at a given time in the statistically stationary regime, the logarithmic representation of $\hat{u}_{\K_{n,a}}(t)$ displays a clear power-law decrease as a function of the radial component when $\rho_i$ remains in the so-called inertial range. Moreover, we observe a rough behavior, as expected given the regularity of a white noise, \emph{i.e.} the independence of the instances as a function of the radial coordinate.

In one spatial dimension $d=1$, we propose the inversion formula \eqref{eq:DeftildeUDeltaPhysSpace} 
\begin{align}\label{eq:DeftildeUDeltaPhysSpace1D}
\widetilde{u}_{\Delta}(t,x) = \sum_{n=1}^N\sum_{m=\pm 1} e^{2i\pi xm\rho_n }\hat{u}_{\K_{n,m}}(t) \Delta \rho ,
\end{align}
with $\rho_n=n\Delta \rho$ and the Hermitian symmetry $\hat{u}_{\K_{n,-1}}(t)=\overline{\hat{u}_{\K_{n,1}}(t)}$ being understood for any integers $1\leqslant  n\leqslant N$.

In Figs. \eqref{fig:Phys1D}(b) and \ref{fig:Phys1D}(c), we display the profiles of the inverse Fourier transform $\widetilde{u}_{\Delta}(t,x)$ \eqref{eq:DeftildeUDeltaPhysSpace1D} as a function of space $x$ at two different viscosities. It is clear that, as viscosity decreases by four orders of magnitude, $\nu=10^{-5}$ in Fig. \eqref{fig:Phys1D}(b)  and $\nu=10^{-9}$ in Fig. \eqref{fig:Phys1D}(c), the solution becomes rougher and rougher, \emph{i.e.} it develops smaller and smaller length scales, as it is expected from the theoretical analysis. Also, we notice that the typical correlation length of the profile is of order of $k_f^{-1}$, as predicted in the theoretical analysis.

\begin{figure}
\begin{center}
\includegraphics[width=8.5cm]{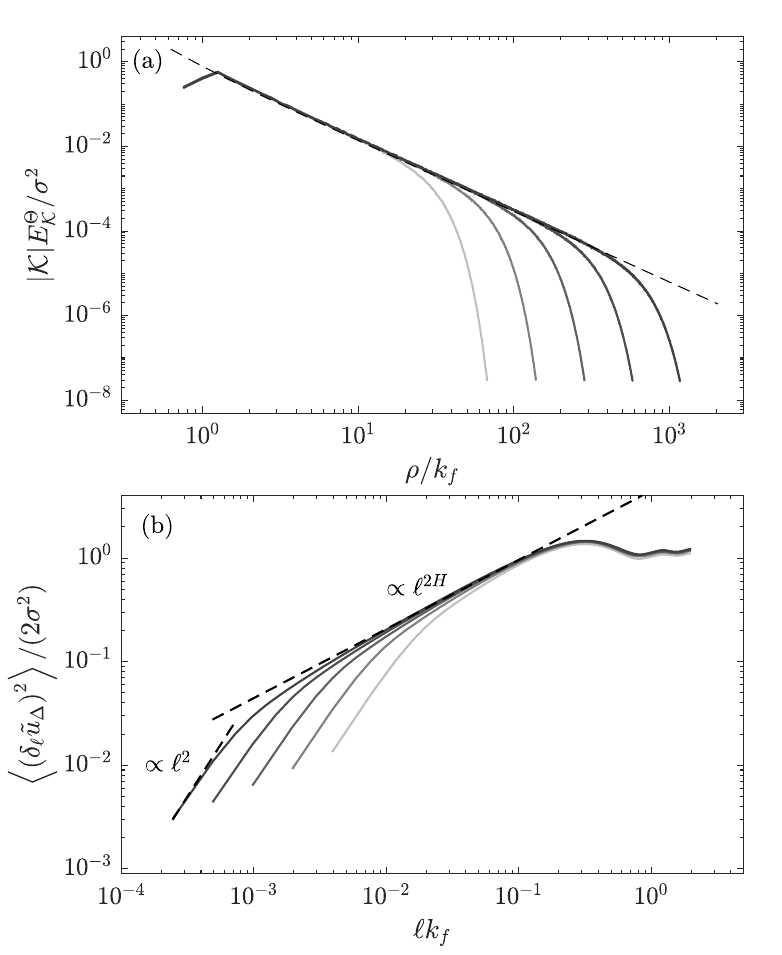}
\end{center}
\caption{\label{fig:Stat1D} Numerical estimation of second-order statistical quantities for the one dimensional $d=1$ fields: (a) Periodograms $E^{\Theta}_\K$ \eqref{eq:DefEKPSDTheta} as a function of the radial coordinate $\rho_i$ and (b) Second-order structure functions $\left\langle \left(\delta_\ell \widetilde{u}_{\Delta}\right)^2\right\rangle$ \eqref{eq:DefS2tildeUDelta}. In both figures, the representation is made in a logarithmic fashion, and the darker the curve, the lower the viscosity. These quantities have been estimated while averaging over $10^3$ instances in the statistically stationary regime, every $10$ time units. All the simulations are conducted with the same parameters $h$, $c$, $H$, $h=\kappa$, $k_f$ and $\widehat{C}_f(k)$ as they are given in the caption of Fig. \eqref{fig:Phys1D}. Values of viscosity correspond to, from lighter to darker, $\nu=10^{-5},10^{-6},10^{-7},10^{-8},10^{-9}$, with corresponding number of collocation points along the radial direction $N=2^{7},2^{8},2^{9},2^{10},2^{11}$. With dashed lines, we superimpose the theoretical predictions of the power-law behaviors in (a) based on \eqref{eq:ExpEKPSDGivenCf} with the particular value $d=1$, and in (b) based on \eqref{eq:ComputSecondOrderSFModelDoubleLimitSmallScales}, with geometrical factor $c_1$ \eqref{eq:PredContCdS2D1}.  }
\end{figure}

Let us now focus on the estimated statistical quantities that have been obtained following the procedure described in paragraph \eqref{Par:AveragProc}. Recall that in dimension $d=1$, averages in the statistically stationary range, are obtained as empirical averages over time, every $10$ time units. In Fig. \eqref{fig:Stat1D}(a), we display the periodogram $E^{\Theta}_\K$ \eqref{eq:DefEKPSDTheta}, based on the finite-volume Fourier mode $\hat{u}_{\K_{n,a}}(t)$, properly weighted by the volume of the unit cell $|\K|$ to make it independent of the resolution $\Delta \rho$, for decreasing values of viscosities (provided in the caption). As we can see, once averaged at various instants of time, which corresponds to the average of the profile represented in Fig. \eqref{fig:Phys1D}(a) at a given viscosity, periodograms are smooth functions of the radial component $\rho_i$, and show at higher wave numbers than $k_f$ and smaller than the characteristic dissipative ones a power-law behavior of exponent $-2H-1$. Recall that we have chosen here the particular value $H=1/3$, thus this power-law decrease, governed by the exponent $-2H-1=-5/3$, corresponds to the one suggested by the phenomenology of fluid turbulence.  As viscosity decreases, the inertial range gets larger and larger, after which the power-law is replaced by an exponential decrease, as it is expected from the action of a viscous Laplacian in the dynamics. 

Let us next derive the precise power-laws that are observed in Fig. \eqref{fig:Stat1D}(a) in order to make our theoretical analysis clear and to check the fine statistical properties of our numerical simulations. This will also allow us to establish a link between the continuous and spectrally finite-volume approaches. We first note that the asymptotic value of the function $F$ \eqref{eq:ComputFDoubleLimit} entering in the expression of the power spectral density \eqref{eq:ComputPSDDoubleLimit} can be easily computed when we take $\widehat{C}_f(k)= \mathds{1}_{\kappa\leq |k| \leqslant k_f}$ (see \eqref{corr-fia}). Indeed, in any spatial dimension  $d$, the PSD \eqref{eq:ComputPSDDoubleLimit} reads
\begin{align}\label{eq:PredPSDContCfNum}
\lim_{\nu\to 0}&\lim_{t\to \infty}E(t,k) =  \frac{\mathds{1}_{\kappa\leq |k| \leqslant k_f}}{2H+d+1}\left[ |k|-\left(\frac{\kappa}{|k|}\right)^{2H+d}\kappa\right]\notag\\
&+ \frac{\mathds{1}_{ |k| \geqslant k_f}}{2H+d+1}\left[ k_f^{2H+d+1}-\kappa^{2H+d+1}\right]|k|^{-(2H+d)}.
\end{align}
Taking into account that the volume of the cell $|\K_{i,a}|$ depends solely on the radial index $i$ and is independent of the angle coordinate $a$, the volume averaged periodogram $E^{\Theta}_\K (t)$  \eqref{eq:DefEKPSDTheta} satisfies
\begin{align}\label{eq:ExpEKPSDGivenCf}
\lim_{\nu\to 0}\lim_{t\to \infty} &E^{\Theta}_\K (t) \notag\\
&\build{\sim}_{\rho_i\to\infty}^{}  \frac{\Delta \rho_i}{|\K_{i,a}|^2}\frac{k_f^{2H+d+1}-\kappa^{2H+d+1}}{2H+d+1}\rho_i^{-(2H+1)}.
\end{align}

We superimpose in Fig. \eqref{fig:Stat1D}(a) with a dashed line the expected asymptotic power-law provided in \eqref{eq:ExpEKPSDGivenCf}, and we observe a perfect matching with the estimates obtained based on our numerical simulations, without any fitting procedure. Notice that, as a consequence of the averaging procedure of the finite volume $\K$, recalling that in dimension $d=1$ the volume $|\K_{i,a}|=\Delta \rho_i=\Delta \rho$ is constant and independent of the index $i$, the periodogram $E_\K$ is expected to be also inversely proportional to the radial discretization $\Delta \rho$, as it is clarified in \eqref{eq:ExpEKPSDGivenCf}. The fact that the periodogram diverges as the volume $\K$ shrinks to 0, i.e. $\Delta \rho\to 0$, is reminiscent of the distributional nature of the continuous solution (see in particular the correlation structure of the continuous modes \eqref{eq:CorrModesSol} that are proportional to a Dirac function).

In a similar fashion to how the periodograms were obtained, we estimate the second order structure function 
$\left\langle \left(\delta_\ell \widetilde{u}_{\Delta}(t,x) \right)^2\right\rangle$ \eqref{eq:DefS2tildeUDelta}, where the expectation is estimated using an empirical average over both time and space, by summing over all computed positions $x$. Once again, our estimates have been obtained in the statistically stationary range and results are expected to be independent of time. We display the results of our simulations and averaging procedure in Fig. \eqref{fig:Stat1D}(b), as a function of the scale $|\ell|$ and for various viscosities. 
At large length scales $|\ell| \geqslant k_f$, \emph{i.e.} above the correlation length scale of the velocity profile $\widetilde{u}_{\Delta}(t,x)$ \eqref{eq:DeftildeUDeltaPhysSpace} in the physical space, the second order structure function  \eqref{eq:DefS2tildeUDelta} reaches a plateau that coincides, in a good approximation, with $2\sigma^2$, where $\sigma^2$ is the variance of the continuous solution \eqref{eq:ComputVarSolDoubleLimit}. Indeed, we could compute in the present situation, with the same $\widehat{C}_f(k)$ as we already specified before, in the limit $\Delta\rho$ much smaller that $k_f$, that, in any spatial dimension $d$,
\begin{align}\label{eq:PredVarContCfNum}
\lim_{\nu\to 0}\sigma^2 = \frac{\Omega_d}{2H(d+1)}\left( k_f^{d+1}-\kappa^{d+1}\right),
\end{align}
which is obtained as the integration over $k\in\R ^d$  of the PSD \eqref{eq:PredPSDContCfNum}. Once non-dimensionalized by $2\sigma^2$, using the formula provided in \eqref{eq:PredVarContCfNum} with the particular value $d=1$ and the relevant values of the other parameters, we can see from inspection of Fig. \eqref{fig:Stat1D}(b) that second-order structure functions reach a plateau of value unity. A power-law behavior of exponent $2H$ follows at smaller scales $|\ell| \leqslant k_f^{-1}$ which remain larger than the dissipative length scale, i.e. in the so-called inertial range of scales, as expected from the asymptotic continuous prediction given in \eqref{eq:ComputSecondOrderSFModelDoubleLimitSmallScales}. Furthermore, we once again observe a perfect match between theory and the results of our simulations when including the multiplicative constant entering in this asymptotic power-law behavior, whose exact expression is also given in \eqref{eq:ComputSecondOrderSFModelDoubleLimitSmallScales}, using in particular the value of the geometric factor $c_1$ \eqref{eq:PredContCdS2D1} predicted in this situation. Finally, in the dissipative range, which is seen at smaller and smaller length scales as the viscosity decreases, we observe the trivial power-law of exponent $2$, which is reminiscent of a smooth behavior, due to the action of the viscous Laplacian in the dynamics.

\begin{figure}
\begin{center}
\includegraphics[width=9.5cm]{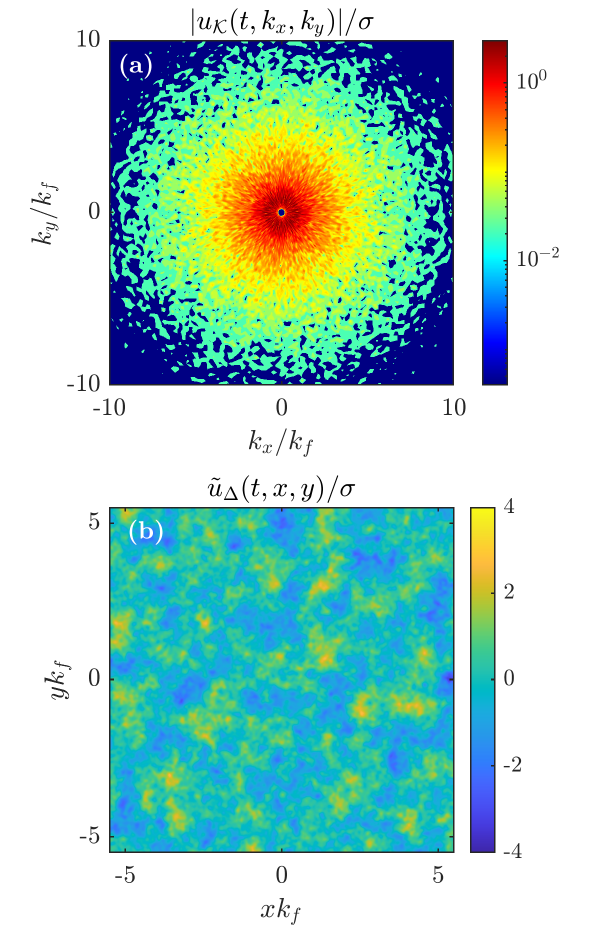}
\end{center}
\caption{\label{fig:Phys2D} Snapshot of the solution for the 2D dynamics, in (a) Fourier and (b) physical spaces, at a time pertaining to the statistically stationary regime. In both cases, we have used the following parameters: $\nu=10^{-5}$, $N=2^{7}$, $H=1/3$, $c=1$, $h=2^{-7}$, $N_\vartheta=2^9$, $\kappa=1$, and $\widehat{C}_f(k)= \mathds{1}_{\kappa\leq |k| \leqslant k_f}$ (see \eqref{corr-fia}), with $k_f=\kappa+3h$.
Notice that to get the inverse Fourier transform $\widetilde{u}_{\Delta}(t,x,y)$, represented in (b), based on the modes $\hat{u}_{\K}(t)$ displayed in (a), we have used the inversion formula based on \eqref{eq:DeftildeUDeltaPhysSpace2D}.  }
\end{figure}

\subsubsection{Two dimensional ($d=2$) simulations}

Let us now present the results of our simulations in spatial dimension $d=2$. In order to do so, we need to discretize also the angle of the polar decomposition used in our finite-volume approach, as detailed in paragraph \ref{Par:DiscreteSd}. Once again, we propagate the solution in time according to the splitting method \eqref{eq:scheme} until time $T^\star$, commented in paragraph \ref{Par:AveragProc}, which can be considered as the beginning of the statistically stationary range.

\begin{figure}
\begin{center}
\includegraphics[width=7.5cm]{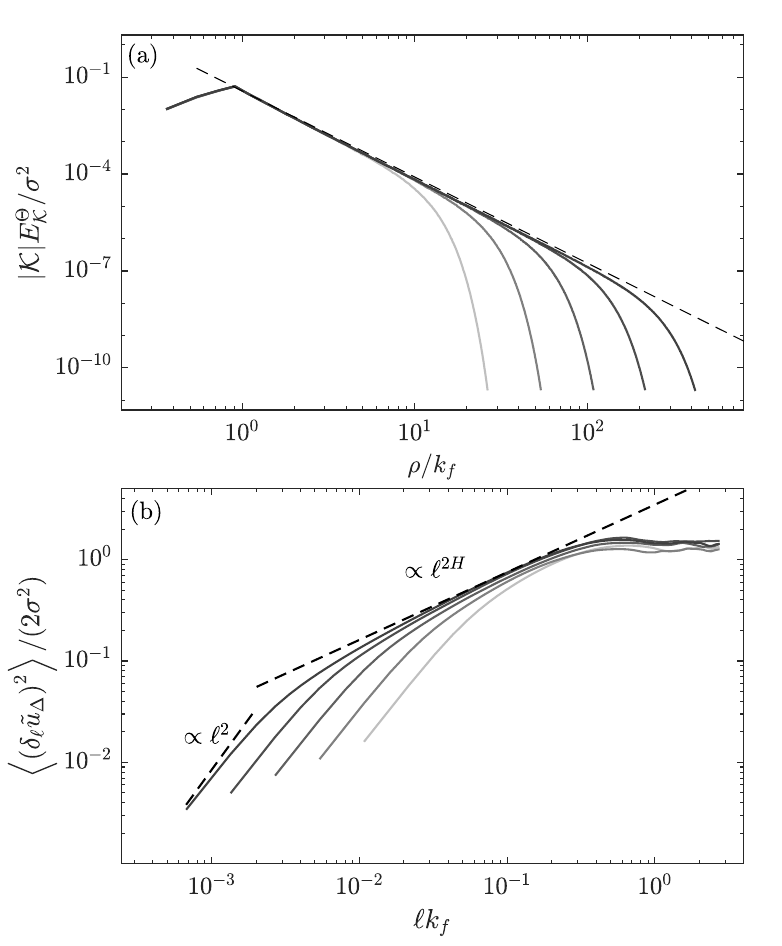}
\end{center}
\caption{\label{fig:Stat2D}  Numerical estimations of the second-order statistical quantities for the two dimensional case ($d=2$). In both figures, darker the curves, lower  viscosities. These quantities are computed in the statistically stationary regime with $500$ instances of the corresponding fields, every  $10$ units of time. All the simulations are conducted with $H=1/3$, $c=1$, $h=2^{-7}$, $N_\vartheta=2^9$, $\kappa=1$, $k_f=\kappa+3h$ and the same $\widehat{C}_f(k)$ used in Fig. \eqref{fig:Phys2D}. Chosen values of viscosities are $\nu=10^{-5},10^{-6},10^{-7},10^{-8},10^{-9}$ with, respectively  $N=2^{10},2^{11},2^{12},2^{13},2^{14}$ collocation points in the radial direction. (a) Angle averaged periodograms $E_\K^\theta$ \eqref{eq:DefEKPSDTheta}, as a function of the radial coordinate $\rho_i$, weighted by the corresponding volume of unit cells $|\K_{i,a}|$ (which is independent of the angle coordinate $a$). (b) Second-order structure functions $\left\langle \left(\delta_\ell \widetilde{u}_{\Delta}(t,x) \right)^2\right\rangle$ \eqref{eq:DefS2tildeUDelta} for different values of the viscosity (solid line). We superimpose with dotted lines in (a) the precise asymptotic power-law behavior given in \eqref{eq:ExpEKPSDGivenCf}, and in (b) the predicted asymptotic power-law based on \eqref{eq:ComputSecondOrderSFModelDoubleLimitSmallScales} with corresponding geometrical factor $c_2$ \eqref{eq:PredContCdS2D2}.  In (b), we also indicate as a guide to the eyes the dissipative behavior $\ell^2$. }
\end{figure}

In Fig. \eqref{fig:Phys2D}(a), we display a logarithmic representation of the finite volume Fourier modes $\hat{u}_{\K}(t)$ in the spectral plane, spanned by the wave vector $k=(k_x,k_y)$, with $k_x=\rho_i\cos \vartheta$ and $k_y=\rho_i\sin\vartheta$. As we can observe, the spectral repartition of energy is clearly isotropic, and it is expected to be statistically invariant by rotation. Also, we notice a rough behavior as a function of the location of the finite volumes, as expected from the independence of the modes, which yields homogeneity in space. This is a clear progress with respect to former numerical approaches that were based on pseudo-spectral simulations, presented in Ref. \cite{ApoBec23}, where strong anisotropies were observed mainly along the horizontal ($\vartheta=0$) and vertical ($\vartheta=\pi/2$) lines. This fully justifies our choice to develop a finite volume approach, beyond the aforementioned theoretical arguments. As we will see later (see Fig. \eqref{fig:Stat2D}), the spectral energy will distribute according to a power-law, and will be ultimately exponentially damped by the action of viscosity.

To get a numerical representation of the counterpart in the physical space of the finite-volume Fourier modes represented in Fig. \eqref{fig:Phys2D}(a), we propose the inversion formula provided in \eqref{eq:DeftildeUDeltaPhysSpace}, which reads explicitly in two-dimensional space $d=2$ as
\begin{align}\label{eq:DeftildeUDeltaPhysSpace2D}
&\widetilde{u}_{\Delta}(t,x,y) = \notag\\
&\sum_{n=1}^N\sum_{m=0}^{2N_\vartheta-1}  e^{2i\pi \left(x\rho_n\cos\vartheta_m+y\rho_n\sin\vartheta_m\right) }\hat{u}_{\K_{n,m}}(t) \rho_n\Delta \rho \Delta \vartheta,
\end{align}
where $\Delta \vartheta = \pi/N_\vartheta$ and $\vartheta_m=m\Delta \vartheta$ (see paragraph \ref{Par:DiscreteSd} for details), and we recall the Hermitian symmetry $\hat{u}_{\K_{n,N_\vartheta+m}}=\overline{\hat{u}_{\K_{n,m}}}$, where $N_\vartheta+m$ is taken modulo $2N_\vartheta$ since $\vartheta$ is defined modulo $2\pi$. Contrary to the one dimensional case ($d=1$), only the parameter $h_i=h$ \eqref{eq:hi-di} is chosen to be a constant, and as a consequence, the radial coordinate $\rho_n$ \eqref{eq:midpointapproximation}, or equivalently the radial stepping $\Delta \rho_n$, of the volumes $\K_{n,a}$ that enters in the expression \eqref{eq:DeftildeUDeltaPhysSpace2D} has to be determined following the recursion procedure specified in \eqref{mesh-generation}. The inversion formula \eqref{eq:DeftildeUDeltaPhysSpace2D} can be further simplified in order to make its numerical computation more efficient. First, let us split the sum over the angular variable into a sum over $m$ between $0$ and $N_\theta-1$ and a second sum for $m$ ranging from $N_\theta$ to $2N_\theta -1$. Shifting the summation variable of $-N_\vartheta$ in the second sum and exploiting the hermitian symmetry, we end up with
\begin{align}\label{eq:DeftildeUDeltaPhysSpace2DHermitian}
&\widetilde{u}_{\Delta}(t,x,y) = \notag\\
&2\Re \sum_{n=1}^N\sum_{m=0}^{N_\vartheta-1} e^{2i\pi \rho_n \left(x\cos\vartheta_m+y\sin\vartheta_m\right) }\hat{u}_{\K_{n,m}}(t) \rho_n\Delta \rho \Delta \vartheta,
\end{align}
where $\Re$ sands for the real part. The transform \eqref{eq:DeftildeUDeltaPhysSpace2DHermitian} is more efficient than \eqref{eq:DeftildeUDeltaPhysSpace2D} from a computational standpoint since it requires two times less operations.

We perform the double series entering in \eqref{eq:DeftildeUDeltaPhysSpace2DHermitian} for each positions $(x,y)\in[-L_{\text{tot}}/2,L_{\text{tot}}/2]$, which are chosen on a uniform cartesian grid, using $\Delta x=1/k_{\text{max}}$ as the spatial resolution in any direction. The largest accessible wave number is given by $k_{\text{max}}=\rho_N$, in a box of length $L_{\text{tot}} = 1/h$. We display the result of this inversion in Fig. \eqref{fig:Phys2D}(b). 

As we can see, the solution $\widetilde{u}_{\Delta}(t,x,y)$ does not exhibit any preferential directions, as it is expected from theoretical predictions and from the statistical homogeneity. Also, the field $\widetilde{u}_{\Delta}(t,x,y)$ is clearly correlated over a finite length scale, which can eventually be related to the characteristic inverse wave number $k_f^{-1}$. Finally, the field seems to develop roughness in the inertial range of scales, and becomes smooth at the smallest scales, i.e. in the dissipative range. This rough behavior will be precisely quantified later while computing respective periodograms and second-order structure functions. Once again, the present numerical method is a real progress compared to pseudo-spectral simulations performed in Ref. \cite{ApoBec23}.

\begin{figure*}
\begin{center}
\includegraphics[width=18cm]{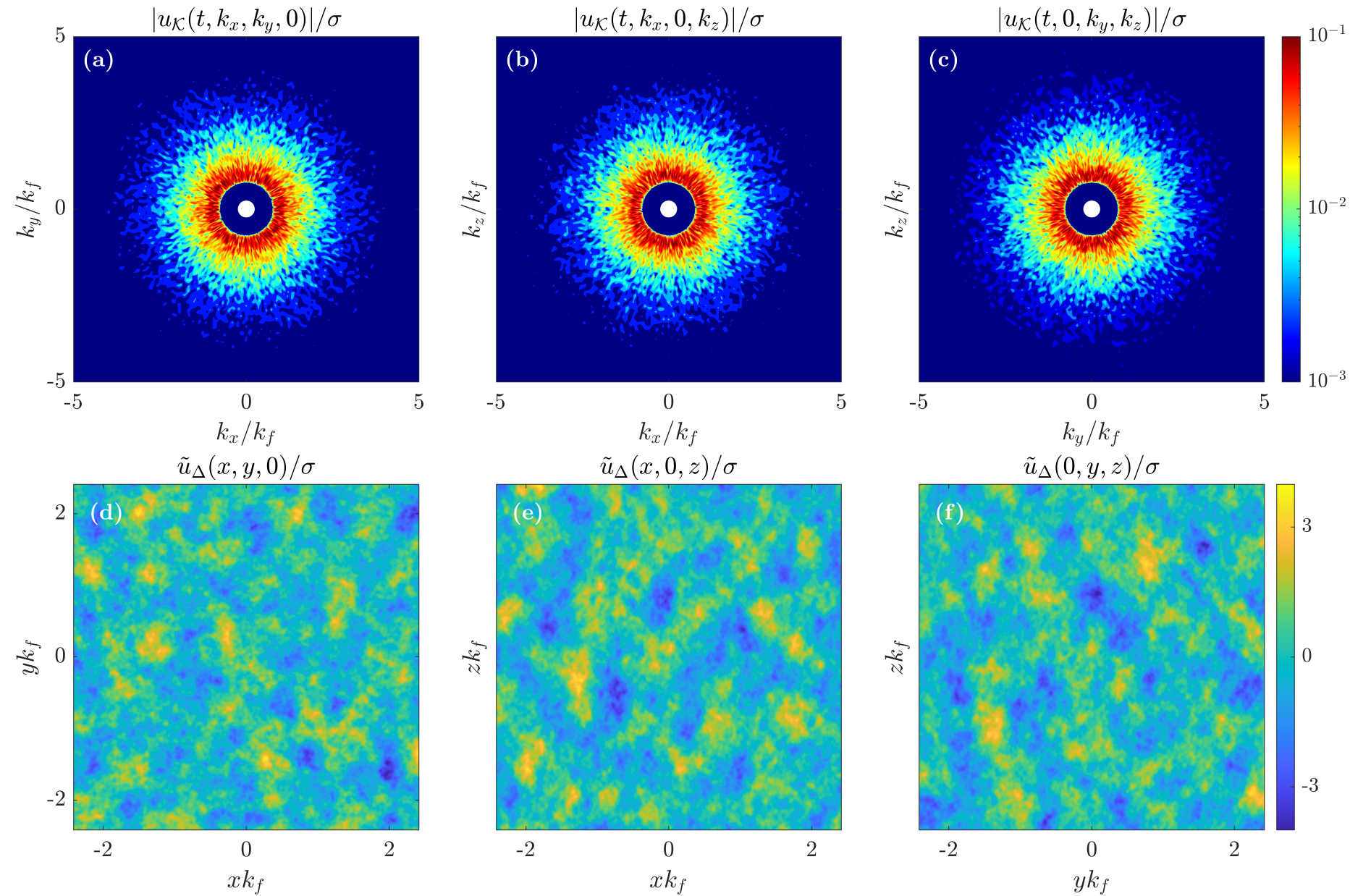}
\end{center}
\caption{\label{fig:Fields3D} Plane cuts of the solution to the 3D Fourier space dynamics. Figures represents the fourier modes in the planes $(k_x,k_y,0)$ (fig (a)), $(k_x,0,k_z)$ (fig (b)) and $(0,k_y,k_z)$  (fig (c)) respectively. Parameters are $H=1/3$, $c=1$, $\kappa=1$, $h=0.15\,\kappa$, $N=2^{8}$, $N_\theta=N_{\varphi}=2^7$, $k_f=\kappa+3h$, $\widehat{C}_f(k)= \mathds{1}_{\kappa\leq |k| \leqslant k_f}$, and $\nu=10^{-6}$. In physical space the cartesian grid is computed over a cubic box of side $L_{\text{tot}}= 1/(2h)$ with a resolution $\Delta x=1/\rho_N $ in every direction.  }
\end{figure*}

Similarly to the one-dimensional case, we now focus on the estimation of the relevant second-oder statistical quantities, including the angle-averaged periodograms $E_\K^\theta$ \eqref{eq:DefEKPSDTheta}. To average in time in the statistically stationary regime, we follow the procedure given in paragraph \eqref{Par:AveragProc}, and we additionally average in this $d=2$ situation over the polar angle $\vartheta$ (see caption of Fig. \eqref{fig:Stat2D} for further details on the statistical sample). We display in Fig. \eqref{fig:Stat2D}(a) the results of our estimations for various viscosities. Notice that in dimension $d=2$, the volume of the cell is given by $|\K_{i,a}|=\rho_i\Delta\rho_i \Delta\vartheta$ which depends on the radial index $i$, thus one must not forget the remaining multiplicative factor $\Delta \rho_i/|\K_{i,a}|^2$ in  \eqref{eq:ExpEKPSDGivenCf}, which has a non trivial dependence on $i$. In particular, because of this factor, $E_\K^\theta$ is not expected to behave as a power-law in the inertial range. As a matter of fact, a power-law behavior is only expected for the quantity $|\K_{i,a}|E^{\Theta}_\K$, as displayed in Fig. \eqref{fig:Stat2D}(a), with an exponent given by $-(2H+2)$. Indeed, in Fig. \eqref{fig:Stat2D}(a) we superimpose the exact formula provided in \eqref{eq:ExpEKPSDGivenCf} and observe that it is in very good agreement with our estimations in the inertial range of scales based on our simulations, without any fitting procedure. Once again, for larger wave numbers in the dissipative range, we observe an exponential decrease.

Let us now display the results for the second-order structure function  $\left\langle \left(\delta_\ell \widetilde{u}_{\Delta}(t,x) \right)^2\right\rangle$ \eqref{eq:DefS2tildeUDelta}, which are estimated using the field $\tilde{u}_\Delta$ defined in \eqref{eq:DeftildeUDeltaPhysSpace2DHermitian}. A value is represented in Fig. \eqref{fig:Phys2D}(b) at a given viscosity $\nu$. Let us recall that we perform an average in time in the statistically stationary regime, and an additional average in space. We display the results of our estimations in Fig. \eqref{fig:Stat2D}(b) as a function of the scale $\ell$, for various values of the viscosities and in a logarithmic representation. Once again, when non-dimensionalized by two times the variance $\sigma^2$ of the solution in physical space, using the formula provided in \eqref{eq:PredVarContCfNum} for the particular case $d=2$, we observe that structure functions at various viscosities reach a plateau of unit value at large scales $\ell \ge k_f^{-1}$. A power-law behavior with the expected exponent $2H$ follows at smaller scales, in the inertial range. This is in a fairly good agreement with the one predicted in \eqref{eq:ComputSecondOrderSFModelDoubleLimitSmallScales}, without any fitting parameter, when taken into account the proper multiplicative factor which includes the geometrical factor $c_2$ \eqref{eq:PredContCdS2D2}. At even smaller scales than the ones of the inertial range, i.e. in the dissipative range where viscosity acts, the smooth behavior is once again recovered with the characteristic power-law exponent $2$.

\subsubsection{Three dimensional ($d=3$) simulations}

We finally explore the instances and statistical behaviors of the solution of our proposed dynamics in space dimension $d=3$. In this situation, when compared with the $d=2$ case, the numerical complexity gets multiplied by the number of discretization points of the second angle $\varphi$ entering the spherical decomposition of the solution $\hat{u}_{\K_{i,a}}$ (see paragraph \ref{Par:DiscreteSd}). A similar remark could be made on a representation of the solution in physical space through an inversion formula \eqref{eq:DeftildeUDeltaPhysSpace}. In this case, this inversion formula reads
\begin{align}\label{eq:DeftildeUDeltaPhysSpace3D}
&\widetilde{u}_{\Delta}(t,x,y,z) = \sum_{n=1}^N\sum_{m=0}^{N_\vartheta-1}\sum_{p=0}^{2N_\varphi-1}\hat{u}_{\K_{n,(m,p)}}(t)\notag\\
&  e^{2i\pi \rho_n\left(x\tilde{k}_{x,(m,p)}+y\tilde{k}_{y,(m,p)}+z\tilde{k}_{z,(m,p)}\right) } \rho_n^2\sin\vartheta_m\Delta \rho \Delta \vartheta\Delta \varphi,
\end{align}
where $\Delta \vartheta= \pi/(N_\vartheta-1)$, $\Delta \varphi= \pi/N_\varphi$ such that $0\leq \vartheta_m=m\Delta \vartheta \leq \pi$ and $0 \leq \varphi_p=p\Delta\varphi < 2\pi$. In \eqref{eq:DeftildeUDeltaPhysSpace3D}, $\Tilde{k}_{(m,p)}=(\tilde{k}_{x,(m,p)},\tilde{k}_{y,(m,p)},\tilde{k}_{z,(m,p)})$ corresponds to the projection of the unit vector in spherical coordinates, that is $\Tilde{k}_{(m,p)}=(\sin\vartheta_m\cos\varphi_p,\sin\vartheta_m\sin\varphi_p,\cos\vartheta_m) $. The hermitian symmetry now writes, $\hat{u}_{\K_{n,(N_{\vartheta} -1 -m,p+N_\varphi)}}=\overline{\hat{u}_{\K_{n,(m,p)}}}$. Similarly to the two dimensional case, $p+N_\varphi$ is taken modulo  $ 2N_\varphi$. This symmetry allows us once again to simplify further \eqref{eq:DeftildeUDeltaPhysSpace3D}. As in the two dimensional case, we split the sum over $p$ into a sum for $p$ between $0$ and $N_\varphi-1$ and a second one for $p$ between $N_\varphi$ and $2N_\varphi-1$. After changing the variables for $l=N_\vartheta-1-m$ and $q=p-N_\varphi$ the second sum writes,
\begin{align}\label{eq:DeftildeUDeltaPhysSpace3DIntermediary}
& \sum_{n=1}^N\sum_{l=N_\vartheta-1}^{0}\sum_{q=0}^{N_\varphi-1} \rho_n^2\sin\vartheta_m\hat{u}_{\K_{n,(N_\vartheta-1-l,q+N_\varphi)}}(t)\notag\\
&  e^{2i\pi \rho_n\left(x\tilde{k}_{x,(N_\vartheta-1-l,q+N_\varphi)}+y\tilde{k}_{y,(N_\vartheta-1-l,q+N_\varphi)}+z\tilde{k}_{z,(N_\vartheta-1-l,q+N_\varphi)}\right) }.
\end{align}
The definition of the angles along with the one of the spherical unit vector $\tilde{k}_{(m,p)}$ yield $\tilde{k}_{(m,p)}=-\tilde{k}_{(N_\vartheta-1-m,p+N_\varphi)}$. This relation combined with the hermitian symmetry and a rearrangement of the sum over $l$ in  \eqref{eq:DeftildeUDeltaPhysSpace3DIntermediary} yields
\begin{align}\label{eq:DeftildeUDeltaPhysSpace3DHermitian}
&\widetilde{u}_{\Delta}(t,x,y,z) = 2\Re\sum_{n=1}^N\sum_{m=0}^{N_\vartheta-1}\sum_{p=0}^{N_\varphi-1}  \hat{u}_{\K_{n,(m,p)}}(t)\notag\\
&  e^{2i\pi \rho_n\left(x\tilde{k}_{x,(m,p)}+y\tilde{k}_{y,(m,p)}+z\tilde{k}_{z,(m,p)}\right) } \rho_n^2\sin\vartheta_m\Delta \rho \Delta \vartheta\Delta \varphi,
\end{align}
Because of the increase of numerical complexity owing to space dimension, we are eventually limited in exploring the behavior of the numerical solution at very low viscosities, which would ask for high values of the radial ($N$) and angular ($N_{\vartheta}$ and $N_{\varphi}$) numbers of collocation points.  We nonetheless managed to perform simulations with reasonable computing power able to represent behaviors expected in an inviscid asymptotic state. In particular, as we will see, our numerical solutions will exhibit power-law behaviors that are characteristic of the asymptotic solutions in the inertial range of scales (see in particular \eqref{eq:ExpEKPSDGivenCf} and \eqref{eq:PredVarContCfNum}).
 \begin{figure}
\begin{center}
\includegraphics[width=8.5cm]{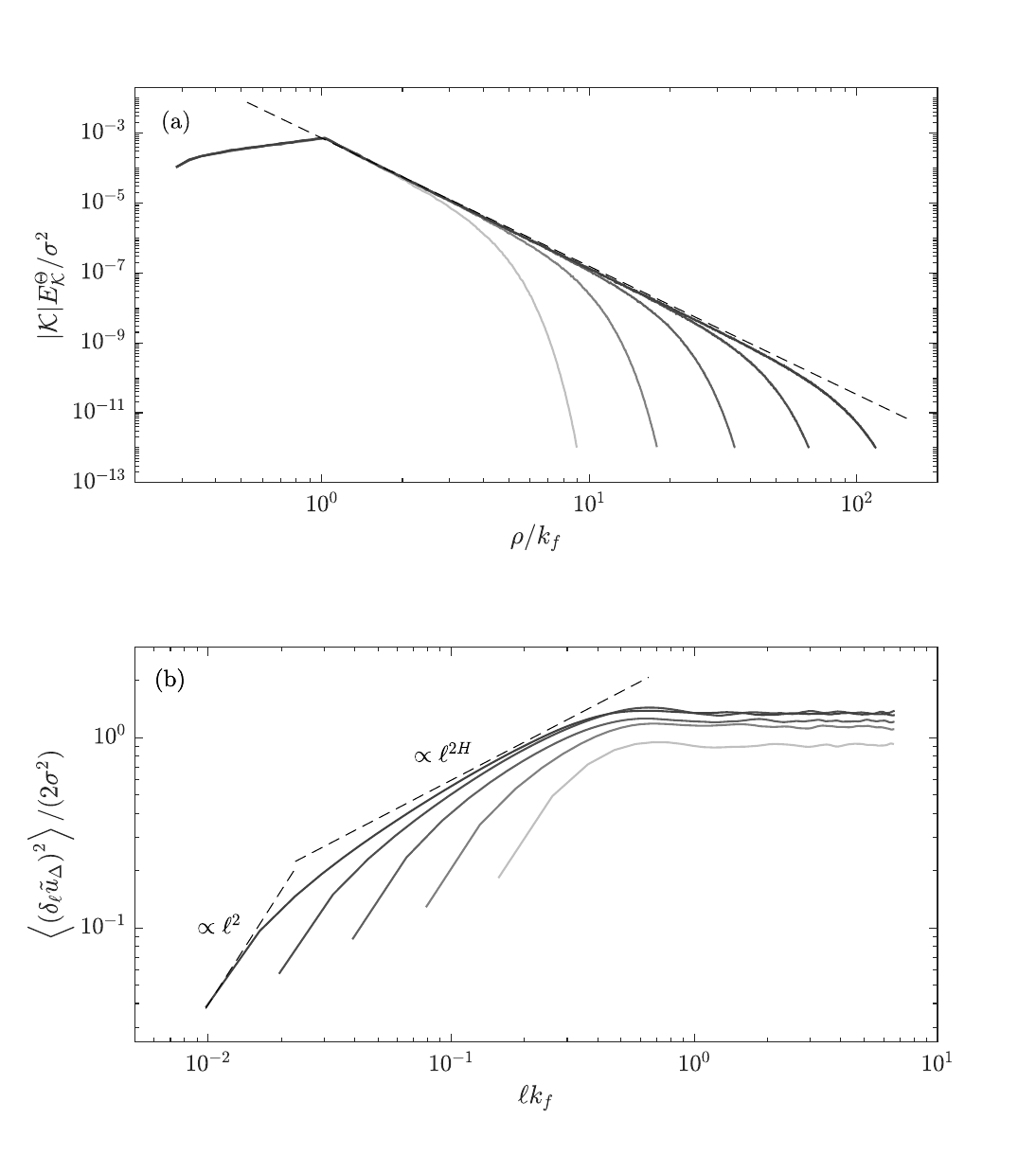}
\end{center}
\caption{\label{fig:Stat3D} Statistical estimation in the three-dimensional case. The statistical sample corresponds to $5$ instances of the fields in time  every $1$ unit of time. The grayscale of each solid lines corresponds to various viscosities. Parameters of the simulations are the same as those provided in the caption of Fig. \eqref{fig:Fields3D}, but for various values of viscosities  $\nu=10^{-5},10^{-6},10^{-7},10^{-8},10^{-9}$ with increasing values of the number of collocation points in the radial coordinate $N=2^{8},2^{9},2^{10},2^{11},2^{12}$. We superimpose with dashed lines the expected asymptotic behaviors in the inertial range, in (a) based on \eqref{eq:ExpEKPSDGivenCf}, and in (b) based on \eqref{eq:ComputSecondOrderSFModelDoubleLimitSmallScales} with the corresponding geometrical factor $c_3$ \eqref{eq:PredContCdS2D3}. In the two cases, we do not make use of additional fitting parameters. In (b), we also indicate the dissipative range with corresponding smooth behavior (i.e. proportional to $\ell^2$).}
\end{figure}

In Figs. \eqref{fig:Fields3D}(a), (b) and (c), we display the  values of the volume-averaged Fourier modes $\hat{u}_{\K_{n,(m,p)}}(t)$ at a given instant in the statistically stationary range and for a given value of viscosity (see caption) in the three planes defined by, respectively, $\tilde{k}_{x,(m,p)}=0$, $\tilde{k}_{y,(m,p)}=0$ and $\tilde{k}_{z,(m,p)}=0$. As we can see, similarly to the $d=2$ case (Fig. \eqref{fig:Phys2D}(a)), the energy is distributed in a statistically isotropic way, i.e. in a rotation invariant way, in all three different planes. Also, notice the rough nature of this distribution, which is reminiscent of the statistical independence of these modes. Once again, this is a real progress compared to pseudo spectral simulations performed in Ref. \cite{ApoBec23} where a strong anisotropy was observed along the axes in the Fourier space.

In Figs. \eqref{fig:Fields3D}(d), (e) and (f), we represent the physical counterpart $\widetilde{u}_{\Delta}(t,x,y,z)$ \eqref{eq:DeftildeUDeltaPhysSpace3DHermitian} of the Fourier modes $\hat{u}_{\K_{i,a}}$ displayed in Figs. \eqref{fig:Fields3D}(a), (b) and (c). These fields are statistically homogeneous and isotropic in a good approximation, with nonetheless some weak anisotropies along the Cartesian directions. We believe that these weak anisotropies are a consequence of the finiteness of radial $\Delta\rho$ and angular $\Delta \vartheta$ and $\Delta \varphi$ steps. We indeed performed other simulations with larger steps (data not shown) which gave stronger anisotropies. Nonetheless, we will see that this spurious anisotropies barely pollute the estimation of forthcoming statistical analyses, that will eventually be in good agreement with theoretical predictions.

As we did at lower space dimensions, we now present the results of the estimation of angular-averaged PSDs and second-order structure functions at various viscosities, and display our results in Fig. \eqref{fig:Stat3D}. We obtain similar results, which include power-law behaviors for $E^{\Theta}_\K$ \eqref{eq:DefEKPSDTheta} (see  Fig. \eqref{fig:Stat3D}(a)) and $\E \left(\delta_\ell \widetilde{u}_{\Delta}(t,x) \right)^2$ \eqref{eq:DefS2tildeUDelta} in the inertial range of scales, corresponding to exponents $-(2H+3)$ (in very good agreement with the analytical prediction provided in \eqref{eq:ExpEKPSDGivenCf} once the nontrivial dependence on the radial coordinate of the volume $|\K_{i,a}|$ of the cells has been taken into account) and $2H$ (also in excellent agreement with the prediction  \eqref{eq:ComputSecondOrderSFModelDoubleLimitSmallScales} with the corresponding geometrical factor $c_3$ \eqref{eq:PredContCdS2D3}), respectively. The action of viscosity can be observed at the highest wave numbers or at the smallest scales. As a final remark, the variance of the fields $\widetilde{u}_{\Delta}$ is very close to the prediction obtained in the continuous framework in the limit of vanishing viscosities \eqref{eq:PredVarContCfNum}. Interestingly, at the highest value of viscosity $\nu=10^{-5}$, we notice that the variance of our numerical simulation underestimate the predicted asymptotic value  \eqref{eq:PredVarContCfNum}. This can be understood as realizing that for such a high value of viscosity, the system is not yet representative of the asymptotic regime $\nu\to 0$ for which the variance is expected independent of $\nu$. This goes the same concerning the power-law behavior in the inertial range. Nonetheless, we can see that as $\nu$ gets smaller and smaller, both variance of the solution, and the power-law behaviors in the inertial range of scales get closer and closer to the asymptotic predictions.

\section{Conclusion and perspectives}\label{Sec:Conclu}

In this article, we have presented an original numerical simulation of a recently proposed dynamics which can be written in Fourier \eqref{eq:DefEvolFourSpace} or in physical \eqref{eq:DefEvolPhysSpace} domains. The underlying physical mechanism is based on a transport of the solution in Fourier space, which must be treated with great care from both a theoretical and numerical points of view. Previous simulations based on pseudo-spectral methods \cite{ApoBec23} were able to give results which are consistent with analytical predictions in a statistical sense, but failed at giving correct solutions. The numerical method proposed in this article is based on a finite-volume approach which allows first to give a proper meaning to the Fourier modes of the solution, and secondly is amenable to effective numerical simulations. Numerical results are in excellent agreement with theoretical predictions, both for fields and for their statistical behaviors.

This investigation leaves many possible extensions and further improvements which are discussed below. 

\subsection{Rigorous analysis of the numerical method}

First, we plan to provide rigorous results on the convergence of the
numerical method proposed in this work. This is challenging because one
needs to identify appropriate functional spaces where the exact and
numerical solutions can be compared quantitatively. Due to the loss of
regularity phenomenon on the considered model, the solutions at finite
times are more regular than at infinite time in the statistically
stationary regime. Being able to state rigorously that the proposed
numerical method reproduces this behavior and in particular reaches a
statistically stationary regime is also an interesting question.

When considered in the Fourier domain, the solution is rough due to the
white noise forcing, which is a non standard and non trivial situation
for finite volume methods, hence the need to develop new tools in the
numerical analysis of the scheme. In addition, the temporal
discretization is based on a splitting method, where solving the radial
transport dynamics exactly is crucial. The condition \eqref{eq:CFL1} is
thus imposed, and it has been explained that it has important
consequences on the geometry of the mesh. In practice, to
increase the stability property of the numerical method or to be able to
introduce other terms in the dynamics, it would be desirable to impose
the more standard and less stringent CFL condition
\[
c \frac{\Delta t}{h_i}\le 1
\]
instead of \eqref{eq:CFL1}. Under this CFL condition, the radial
transport dynamics is not solved exactly anymore and it is well-known
that numerical dissipation and regularization effects appear \cite{Lev92,DelLag11,JunPer23}. In that
setting, it is not clear whether the numerical method would be able to
reproduce the qualitative behavior of the model.

Finally, note that we are mainly interested in the approximation of the
probability distribution of the solution. This calls for the application
of weak error analysis techniques to obtain error estimates, which need
to be developed for the considered model.

\subsection{Generalization to vector  fields}

From a physical point of view, whereas it was of tremendous importance to design proper generalizations to higher dimensions, as we did in a theoretical fashion in Ref. \cite{ApoBec23}, we now need to deal with vector fields, in particular incompressible ones (i.e. divergence-free) if we want to propose realistic models of fluid mechanics in turbulent situations. Some progress have already been done in that direction and will be communicated in the near future. 

\subsection{Definition and computation of the inverse Fourier transform}

We also need to give a proper meaning to the physical-space counterpart of the finite-volume spectral field $\hat{u}(t,k)$ \eqref{eq:DefGeneuK}. We have here proposed an approximation $\widetilde{u}_{\Delta}(t,x)$ of such a field based on the inversion formula \eqref{eq:DeftildeUDeltaPhysSpace}. It would be insightful to make a clear link between this approximation and the solution $u(t,x)$ of the continuous formulation \eqref{eq:DefEvolPhysSpace}.

From a physical viewpoint, obtaining the solution in physical space allows to answer several questions regarding fluid mechanics. Indeed, the present approach focuses on the Eulerian framework, that is obtaining a modeled fiels which depends on space and time. Another description of fluids, called the Lagrangian formulation, focuses on the \textit{flow} of velocity fields, i.e. the velocity of tagged fluid particules along their trajectory $X(t)$. The path $X(t)$ is defined as the solution of the flow equation $dX/dt=u(t,X)$ given an initial position $X(t_0)$, in a unique fashion if the advecting field is smooth in space, while $u$ has been possibly generalized to a divergence-free vector valued field. The Lagrangian investigation of laboratory and numerical turbulent flows has been intensively developed over the last thirty years, as reviewed in \cite{Yeu02,TosBod09} and \cite{PinSaw12}, following an intense and vast effort aimed at characterizing with precision the statistical behavior of the Eulerian velocity field \cite{Fri95}. An important question would be devoted to the consequence on the regularity of the velocity $v(t)=dX/dt$ along trajectories while imposing a given Hurst exponent $H$ on the Eulerian field $u(t,x)$, as it has been preliminary explored in a different setup in \cite{RenChe20}.

\subsection{Modeling the intermittency phenomenon}

Also, we have here focused at modeling fluid turbulence at a statistical level up second-order. Observations based on experimental and numerical investigations of the turbulent velocity field show that velocity fluctuations are non-Gaussian, which is usually refereed in turbulence literature as the intermittency phenomenon \cite{Fri95}. In the present approach, we could wonder how to include the intrinsically non-Gaussian nature of the fluctuations at the finest scales. A first proposition was made in Ref. \cite{ApoChe22} consisting in defining the particular case $H=0$, that leads to logarithmically correlated Gaussian fields, corresponding to the Gaussian free field in space dimension $d=2$ \cite{She07}, which are known when exponentiated to lead to a Gaussian multiplicative chaos measure \cite{Kah85,RhoVar14}. Additional heuristics were then provided in Ref. \cite{ApoChe22} to include such a probabilistic object in a dynamical picture, leading to nonlinear (quadratic) corrections to the evolution of the type proposed in \eqref{eq:DefEvolPhysSpace}, in a different, but  complementary, spirit than the approach developed in Ref. \cite{BanEce23}. The analysis of such a nonlinear evolution gets much more complicated and we could wonder whether tractable approaches could be designed amenable to rigorous treatments. We also keep these developments for future investigations.

\subsection{Kinetic energy budget and limitations of the present dynamical model regarding the phenomenology of turbulence}

In the present evolution \eqref{eq:DefEvolPhysSpace}, we could wonder about identifying the mechanism that plays a key role in the establishment of a statistically stationary regime, in which the velocity variance gets finite, and moreover uniformly bounded with viscosity \eqref{eq:ComputVarSolDoubleLimit}. To do so, we need to derive the kinetic energy budget, that is the time evolution of the velocity variance $\E \left( u(t,x) \right)^2$. Concerning real fluids, governed by the Navier-Stokes equations \eqref{eq:NS}, this budget is crucial in the understanding of the core of turbulent phenomenology, can be derived locally (i.e. without expectations), as it is done in Ref. \cite{DucRob00} and reviewed in Refs. \cite{EyiSre06,Eyi07} in the context of the Onsager's conjecture. In this case, assuming incompressibility, and statistical homogeneity and isotropy, it can be derive from \eqref{eq:NS} in a straightforward manner that the kinetic energy budget which governs the time variation $d\E \left| u(t,x) \right|^2/dt$ is solely given by the competition between energy injection and viscous diffusion. Concerning the present model, with evolution in physical space given in \eqref{eq:DefEvolPhysSpace}, this balance is modified by the transport operator $A$ entering in the evolution. 

Since the solution at any time $t$ is statistically homogeneous (see in particular the formal solution in Fourier space \eqref{eq:MildSolution}), taking into account the independence in time of the instances of the forcing \eqref{eq:CorrForcingPhysSpace} that asks for applying It\^o's lemma, we obtain from \eqref{eq:DefEvolPhysSpace} in an exact manner that
\begin{align}\label{eq:KEBmodel}
\frac{1}{2}\frac{d}{dt}\E \left( u\right)^2 = \frac{1}{2}C_f(0) - \nu \E\left| \nabla u\right|^2 - \E\left( uAu\right),
\end{align}
where it is understood that the solution $u$ is evaluated at time $t$ and at the position $x$. The first term entering at the RHS of the kinetic energy budget \eqref{eq:KEBmodel} is the energy injection, which form is a consequence of  It\^o's lemma, and $C_f (0)$ is the value at the origin of the spatial correlation structure of the forcing $f$ \eqref{eq:CorrForcingPhysSpace}. The second term, obtained from $ \E\left( u \Delta u\right)$ after having used the statistical homogeneity of the solution, represents viscous dissipation, and the last term quantifies the contribution of the operator $A$ to this budget. As time goes on, we have already shown that the solution reaches a statistically stationary regime in which the variance of the solution gets finite and independent of viscosity as $\nu\to 0$ \eqref{eq:ComputVarSolDoubleLimit}. The question that we want to address now is what are the role played by the terms entering in the RHS of \eqref{eq:KEBmodel} to ensure that $d\E \left( u\right)^2/dt\to 0 $ as $t\to \infty$.

Let us focus on the contribution of the operator $A$ in the kinetic budget (last term on the RHS of \eqref{eq:KEBmodel}). Making use of the Parseval's identity, and the formal solution of the evolution provided in  \eqref{eq:MildSolution}, which in particular leads to the independence of the Fourier modes \eqref{eq:CorrModesSol}, we can easily see that the transport term (expressed as a divergence with respect to the wave vector $k$ that enters in the operator \eqref{eq:DefAPseudoDiffPhysSpace}) does not contribute, only the additional dissipation term proportional to $|k|^{-1}$ eventually contributes. Indeed, making use of \eqref{eq:FnuFiniteTime}, we can then easily get that
\begin{align}
\E\left( uAu\right)&=\notag\\
&c\left(H+\frac{1}{2}\right)\int |k|^{-(2H+d+1)} e^{-\frac{8\pi^2 \nu}{3c} |k|^3} F_\nu(t,|k|)dk, \notag
\end{align}
such that at infinite time and as viscosity goes to $0$, using the respective expression $F(|k|)$ of $F_\nu(t,|k|)$  in this double limit \eqref{eq:ComputFDoubleLimit}, we obtain
\begin{align}
\lim_{\nu\to 0}\lim_{t\to \infty}\E\left( uAu\right)= \frac{1}{2}C_f(0),\notag
\end{align}
which says that the contribution of the operator $A$ in the kinetic energy energy budget \eqref{eq:KEBmodel} compensates in an exact manner the energy injection in this double limit. As consequence, because the LHS of \eqref{eq:KEBmodel} vanishes at infinite time for any values of viscosity, this means that viscous dissipation has no contribution as $\nu\to 0$. This behavior differs from the one observed for the Navier-Stokes where it is expected that velocity gradients variance diverges exactly as the inverse of viscosity \cite{Fri95,Pop00,TenLum72,EyiSre06,Dub19}. Instead, we can show, using the formal solution provided in \eqref{eq:MildSolution} that velocity gradient variance diverges much slower than $\nu^{-1}$ as $\nu\to 0$, according to 
\begin{align}
0<\lim_{\nu\to 0}\lim_{t\to \infty} \nu^{\frac{2}{3}(1-H)} \E\left| \nabla u\right|^2 <+\infty, \notag
\end{align}
similarly to what was found in the simpler framework developed in Ref. \cite{ApoChe22}. 

Thus, it would be of tremendous importance from a physical point of view to build a simple dynamics able to reproduce the two aforementioned properties of fluid turbulence, that is the finiteness of velocity variance \eqref{eq:AsymptVar} and the generation of small scales according to the Kolmogorov spectrum \eqref{eq:Kolmo41}, while ensuring the independence of the average dissipation $\nu \E\left| \nabla u\right|^2$  with respect to viscosity as the latter gets smaller and smaller. We keep this perspective for future investigations.

\begin{acknowledgments}
We would like to thank Isabelle Gallagher, Fr\'ederic Lagouti\`ere and Antoine Mouzard for many discussions concerning this matter. LC is partly supported by the Simons Foundation Award number 1151711.  CEB is partially supported by the project SIMALIN (ANR-19-CE40-0016) operated by the French National Research Agency.
\end{acknowledgments}

%%%%%%%%%%%%

%\bibliographystyle{plain}
%\bibliographystyle{unsrt}
%\bibliography{/Users/lchevill/ownCloud/Redac/MyBibTex/mybiblioJune2015}
%\bibliography{mybiblioJune2015}
%\bibliography{/home/lchevill/Redac/MyBibTex/mybiblioJune2015}

\end{document}